\documentclass[aps,prd,tighten,amssymb,epsfig,showpacs,twocolumn]{revtex4}
\usepackage{epsfig}
\usepackage{amsmath}
\usepackage{mathrsfs}
        
\newcommand{\scri}{\mathscr{I}}

\begin{document}
\title{Learning about compact binary merger: the interplay between 
numerical relativity and gravitational-wave astronomy}

\author{Thomas Baumgarte${}^{1,2}$,
	Patrick Brady${}^{3}$,
        Jolien D E Creighton${}^{3}$,
	Luis Lehner${}^{4}$,
	Frans Pretorius${}^{5,6,7}$,
        Ricky DeVoe${}^{8}$}

\affiliation{
${}^{1}$Department of Physics and Astronomy, Bowdoin College, Brunswick,
ME 04011 \\	
${}^{2}$Department of Physics, University of Illinois at Urbana-Champaign,
Urbana, Il 61801 \\
${}^{3}$Department Physics,
University of Wisconsin--Milwaukee, P.O. Box 413, Milwaukee, WI 53201\\
${}^{4}$Department of Physics and Astronomy,
Louisiana State University, Baton Rouge, LA 70810, \\
${}^{5}$Department of Physics, University of Alberta, Edmonton, AB T6G 2G7,\\
${}^{6}$Canadian Institute for Advanced Research, Cosmology and Gravity Program\\
${}^{7}$Department of Physics, Princeton University, Princeton, NJ 08540\\
${}^{8}$Beloit College, 700 College St., Beloit, WI 53511 
}

\begin{abstract}
Activities in data analysis and numerical simulation of
gravitational waves have to date largely proceeded independently.
In this work we study how waveforms obtained from numerical simulations 
could be effectively used within the data analysis effort to search
for gravitational waves from black hole binaries.
We propose measures to quantify the accuracy of numerical
waveforms for the purpose of data analysis and study how sensitive the 
analysis is to errors in the waveforms. We estimate that $\sim 100$
templates (and $\sim 10$ simulations with different mass ratios) are
needed to detect waves from \emph{non-spinning} binary black holes
with total masses in the range $100 M_\odot \leq M \leq 400 M_\odot$
using initial LIGO.
Of course, many more simulation runs will be needed to confirm that the
correct physics is captured in the numerical evolutions.  From this
perspective, we also discuss sources of systematic errors in numerical waveform 
extraction and provide order of magnitude estimates for the computational
cost of simulations that could be used to estimate the cost of 
parameter space surveys.  Finally, we discuss what information from 
near-future numerical simulations of compact binary systems would be 
most useful for enhancing the detectability of such events with contemporary
gravitational wave detectors and emphasize the role of numerical
simulations for the interpretation of eventual gravitational-wave
observations.
\end{abstract}

\maketitle

\section{Introduction}
\label{s:intro}

\begin{figure}
\includegraphics[width=3.6in,clip]{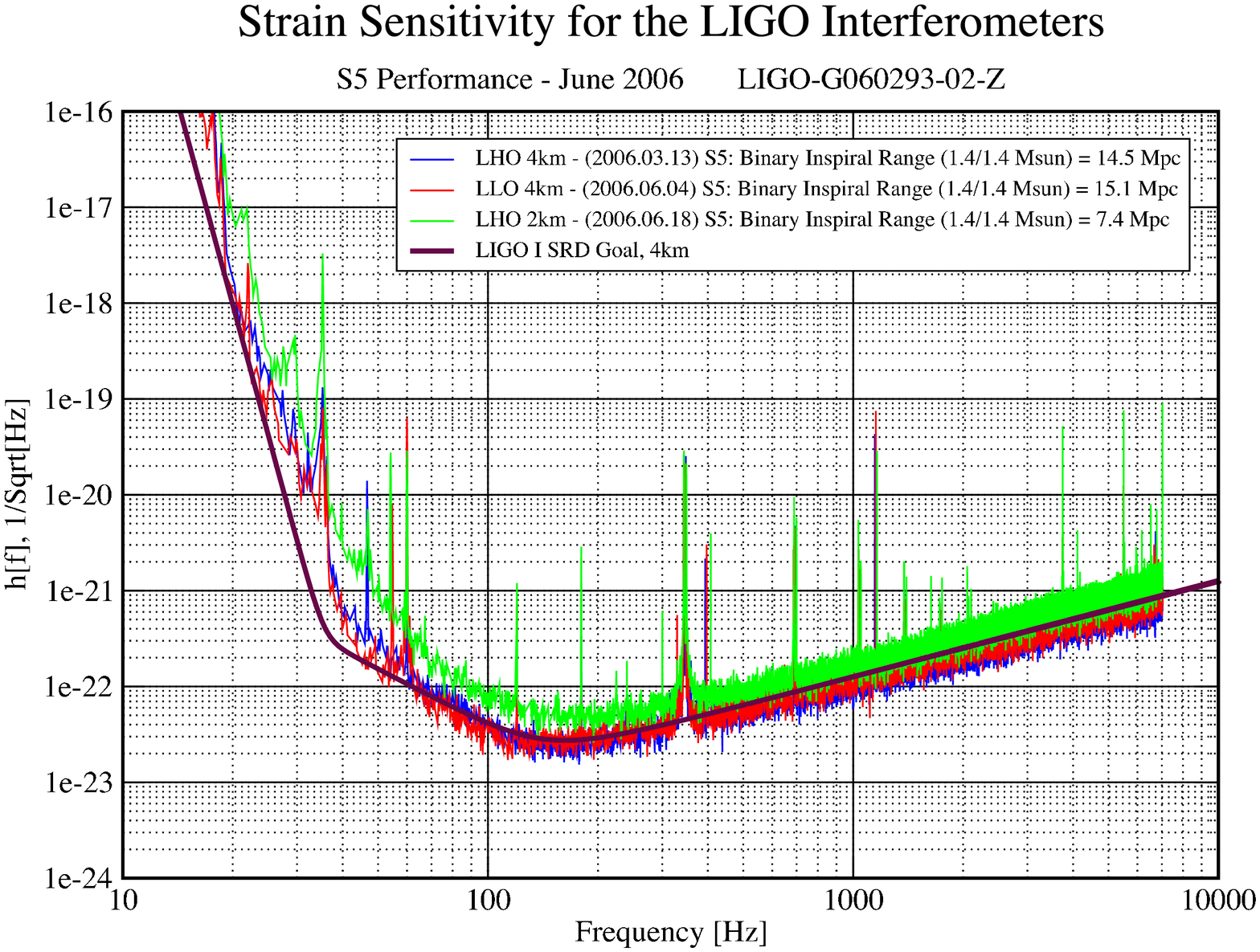}
\caption{\label{f:ligo-noise}
The noise sensitivity curves for the LIGO interferometers published in
June 2006~\cite{ligo-noise}.  The blue and red curves are the 4km
interferometers at Hanford and Livingston, respectively.  The green
curve is the 2km Hanford interferometer.  The LIGO-I noise curve used
for the sample calculations in this paper is the solid purple line.
Compact binaries generate gravitational-waves which sweep upward in
frequency as they inspiral and merge.  The frequency ($\approx 40
\textrm{ Hz}$) below which the noise curve rises sharply determines
the longest dynamical time-scale of the sources to which the LIGO
instruments are sensitive; this, in turn, translates to a largest mass
compact binary system to which LIGO is sensitive.
}
\end{figure}

Searches for gravitational waves from coalescing compact binary
systems rely on concrete knowledge of the waveform to achieve maximum
sensitivity to these sources. With LIGO currently acquiring data at
design sensitivity (see Fig.~\ref{f:ligo-noise}), an optimal matched filtering search could 
detect the gravitational waves from binary black hole coalescence out 
to several hundred Mpc. Direct observation of
gravitational waves from these systems will have significant and
far reaching consequences for both gravitational physics and astronomy.

To date, searches for gravitational waves from compact binary systems
using data taken at LIGO, GEO and TAMA observatories have concentrated
mostly on binary neutron stars and speculative lower mass
systems~\cite{gwsearches} -- each element of the binary has a mass
below $m_j \leq 3 M_{\odot}$.  Searches for inspiral waves from higher
mass systems such as binary black holes and black-hole neutron star
pairs have used detection templates constructed to match with a wide
variety of theoretical waveforms~\cite{Abbott:2005kq}.  This
is the first step in searching for one of the most promising and
tantalizing sources accessible to earth-based gravitational wave
detectors.  As numerical relativity simulations produce waveforms, new
issues arise when trying to migrate this knowledge into the analysis
efforts.

The gravitational waves measured from a compact binary system depend
on a number of parameters including the masses $m_1$ and $m_2$, the
spins $\vec{s}_1$ and $\vec{s}_2$, the time of merger $t_0$, the
inclination of the orbital plane $\iota$, the phase of the orbit at
time of merger $\Phi_0$, the distance to the binary $D$, the location
of the binary on the sky and the polarization angle between the
propagation and detector frame.  For binary black holes these are all
the free parameters; for neutron stars there are additional parameters
that relate to their internal structure and composition.  In
considering the utility of numerical simulations in gravitational-wave
astronomy, it is critical to understand the dependence of the
numerical waveforms on all of these parameters.  Some of the
dependencies are, in principle, in hand already.  For example, the
functional form of the dependence on sky location, polarization angle
and distance is known analytically if the waves can be extracted
accurately from numerical simulations. The time of merger is also easily
accounted for by translating the computed waveform in time (which is
done by applying a frequency dependent phase shift to the waveform).

Post-Newtonian calculations of the waveforms demonstrate the
complicated dependence on the other parameters. When all known
amplitude and phase terms are included in the waveform, it is
necessary to explicitly measure the masses, spins, inclination and
phase. On the other hand, the so-called restricted post-Newtonian
templates (which keep all phase corrections but only the leading amplitude
term) provide a simplified and efficient detection method for
non-spinning binaries. The data analysis problem (for fixed masses)
reduces to the detection of a signal of unknown amplitude and phase
for which the optimal method is well known~\cite{weinsteinZubakov}.

In this note, we present preliminary investigations of waveform
accuracy and discuss implications for numerical simulations. 
In Sec. \ref{s:waveforms}, we give an overview of the form of the
gravitational wave strain that is measured at detectors, the relationship
between this and a common way of describing the waveform 
from numerical simulations in terms of Newman-Penrose scalars, and
summarize the numerical results we will use as examples.
In Sec. \ref{s:accuracy}, we
provide a framework to obtain an estimate of the accuracy of a
particular simulation with relation to the data analysis
techniques. We present examples from the perspective of both
inspiral and burst searches, and from the inspiral study
we obtain a crude estimate that, for the purposes of {\em detection} of
{\em non-spinning} binaries, a template bank for initial LIGO 
containing approximately 100 waveforms will be sufficient; this
corresponds to about 10 simulations with different mass ratios.
In Sec. \ref{s:discussion}, we conclude with broad discussions
of several important and related issues, including
certain technical aspects of waveform extraction from
simulations that might introduce spurious effects 
---details of this discussion are deferred to Appendix \ref{appwaves}. 
We also explain how information from current simulations, before
fully qualified simulations can produce template waveforms, can be
used to enhance the detection of binary black holes using excess power
type searches tuned to the right frequencies. Finally, we discuss how
numerical simulations may be used to interpret eventual
gravitational-wave observations. Most of our analysis focuses on binary
black hole systems, though in Sec. \ref{s:discussion} we also comment on 
other compact binaries that contain one or two neutron stars.  
In Appendix \ref{app_cost} we provide scaling estimates of the CPU time 
(cost) required to simulate binary systems. This should be
useful to estimate how feasible various surveys of binary
black hole merger parameter space would be on contemporary
computer systems.\\

\section{Gravitational waveforms from black hole binaries}
\label{s:waveforms}

For a binary black hole system, the gravitational-wave strain measured
at one of the detectors can be written as
\begin{widetext}
\begin{equation}
  h(t) = \frac{1 \textrm{Mpc}}{D} \left\{
  F_+(\alpha , \delta, t, \psi) 
  h_+( t-t_0; m_1, m_2, \vec{s}_1, \vec{s}_2, \iota, \Phi_0) +
  F_\times(\alpha , \delta, t, \psi) 
  h_\times( t-t_0; m_1, m_2, \vec{s}_1, \vec{s}_2, \iota, \Phi_0)
  \right\} 
  \label{e:detector-strain}
\end{equation}
\end{widetext}
where $m_1$ and $m_2$ are the black hole masses, 
$\vec{s}_1$ and $\vec{s}_2$ are the black hole spins, 
$t_0$ is the time of merger, the inclination of the orbital
plane is $\iota$, the phase of the orbit at time of merger is
$\Phi_0$, the distance to the binary is $D$ (measured in Mpc) and
$(\alpha, \delta)$ are the right ascension and declination of the binary 
and $\psi$ is the polarization angle between the propagation and
detector frame. 

Numerical waveforms are usually obtained from the Newman-Penrose
coefficient $\Psi_4$.  Under a careful choice of coordinates,
frame and extraction world tube (further discussion of which 
is presented in Sec.\ref{s:discussion} and Appendix \ref{appwaves}) the 
waveform is related to $\Psi_4$ by
\begin{equation}\label{Psi4_hpc_def}
\frac{d^2}{dt^2} (h_+ + i h_\times) = \Psi_4 \; .
\end{equation}
It is often convenient to represent the waveform in terms of
spin-weight $-2$ spherical harmonics as
\begin{equation}
h_{+,\times} = \sum c_{+,\times}^{lm} {}_{-2}Y_{lm} \; .
\end{equation}
For concreteness, we consider waveforms extracted from numerical
simulations performed by Pretorius~\cite{Pretorius:2005gq}; similar
calculations can certainly be done with waveforms extracted from
other
simulations~\cite{Bakeretal1,CLMZ,HLS,S,gonzalez_et_al,bruegmann_et_al}.

\begin{figure}
\includegraphics[width=3.3in,clip]{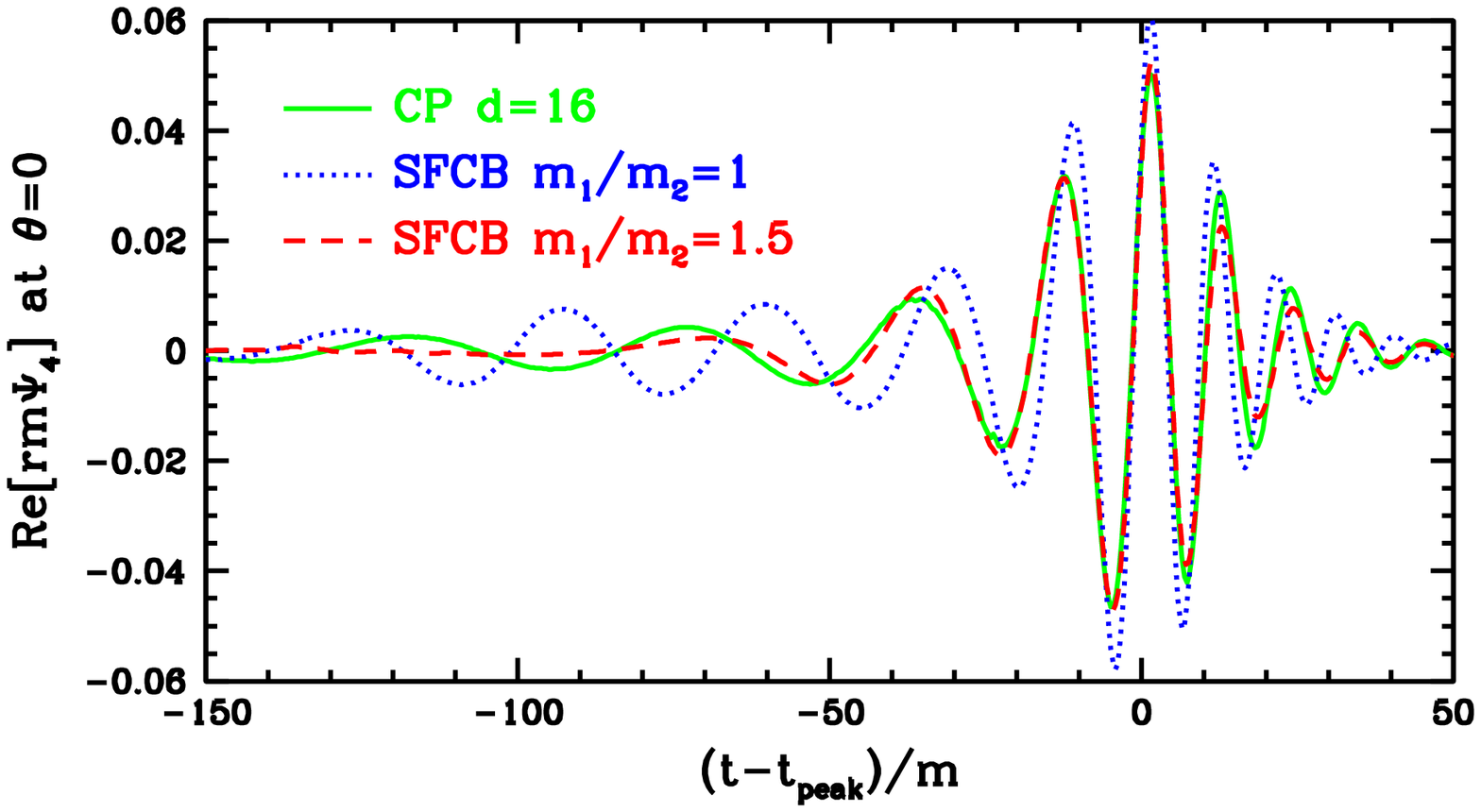}
\includegraphics[width=3.3in,clip]{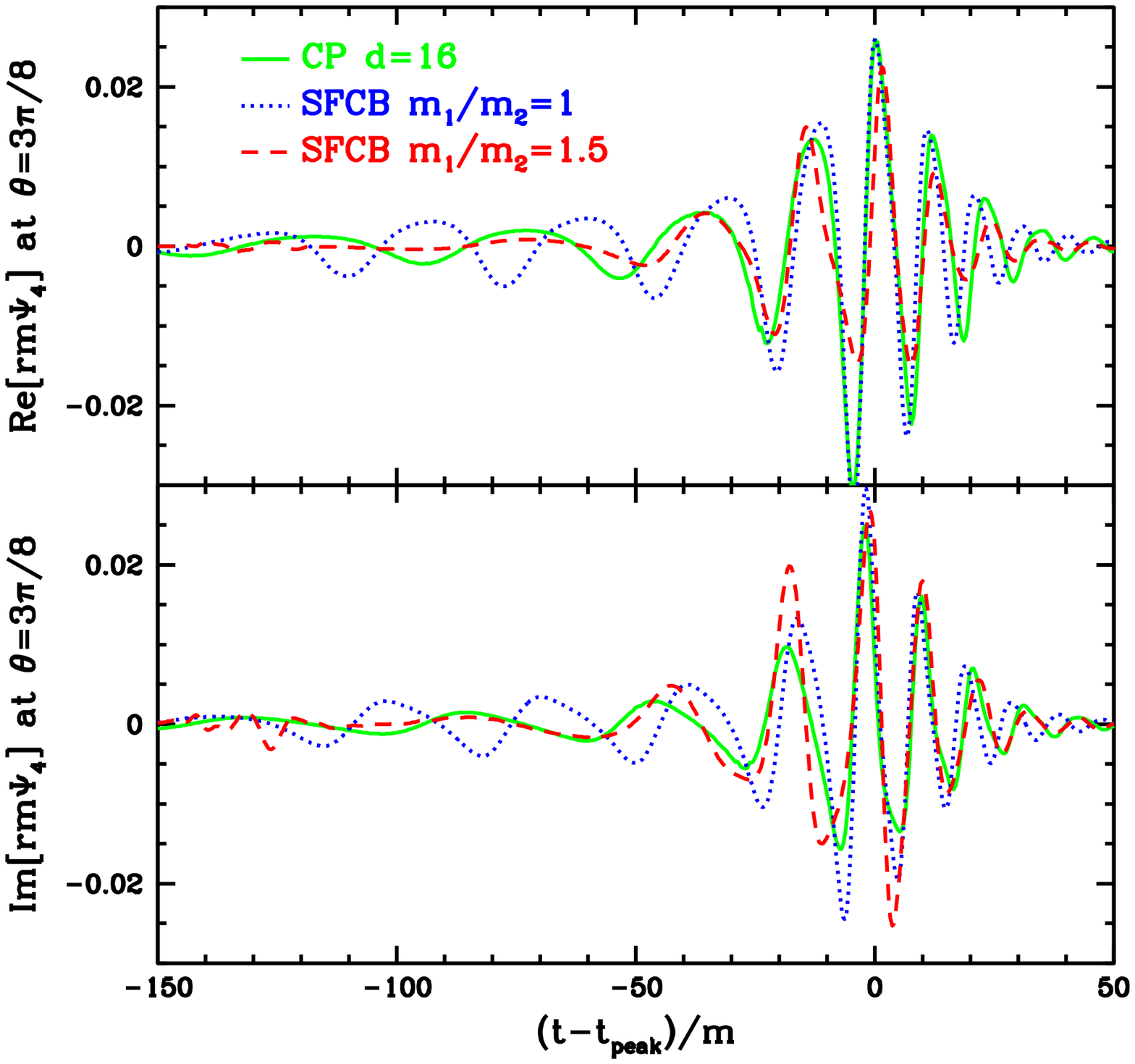}
\caption{\label{f:waveforms}
Samples of $\Psi_4$ from the binary black hole merger simulations discussed 
here. Evolutions
from three different initial conditions are shown: Cook-Pfeiffer $d=16$ (CP d=16),
and two scalar field collapse binaries (SFCB), one equal mass, the other
with a mass ratio of $1.5:1$.
The top plot shows the real part of $\Psi_4$ evaluated along the
axis $\theta=0$ orthogonal to the orbital plane (and azimuthal
angle $\phi=0$); for brevity
we do not show the imaginary part as it looks almost identical
modulo a phase shift. The figures below show the real and
imaginary parts of $\Psi_4$ evaluated at $\theta=3\pi/8$ (note
the different vertical scale). Here we show both components as
there are noticeable differences between the two polarizations.
In all cases the waveform was extracted at a coordinate
radius of $r=50m$, where $m$ is the sum of initial apparent horizon
masses; also, the time has been shifted so that $t=0$ corresponds
to the peak in wave amplitude, and $\Psi_4$ has been multiplied
by a constant complex phase angle to aid comparison.}
\end{figure}

\begin{figure}
\includegraphics[width=3.3in,clip]{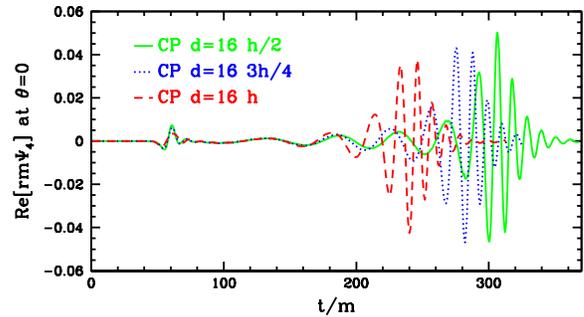}
\caption{\label{f:cnv_waveforms}
A plot demonstrating the dependence on numerical resolution
of Cook-Pfeiffer $d=16$ initial data evolutions. The lowest characteristic
resolution (dashed line) has a characteristic mesh spacing of $h$, the next
lowest one of $3h/4$ (dotted) while the finest resolution
has a mesh spacing of $h/2$ (solid). The dominant
component of the numerical error is in the phase
evolution of the inspiral portion of the wave.
See~\cite{buonanno_et_al}
for a detailed discussion of the numerical errors in this
set of evolutions.}
\end{figure}

The simulations evolve two classes of initial data, a) equal mass,
quasi-circular co-rotating initial configurations as calculated by
Cook and Pfeiffer~\cite{CP,CPID,PKST}, and b) black hole binary
systems formed via the gravitational collapse of two boosted scalar
field pulses. A detailed analysis of the Cook-Pfeiffer (CP) evolutions is
presented in~\cite{buonanno_et_al}.  The scalar field collapse
binaries (SFCB) have non-negligible initial eccentricity, zero initial spin, and
simulations with mass ratios up to $1.5:1$ have been performed (a
more detailed description of the equal mass scenarios can be found
in~\cite{Pretorius:2005gq,pretoriusmomentum}). Note that the scalar
field is merely a convenient vehicle to create binary configurations;
remnant scalar field energy leaves the vicinity of the binaries in
about one light crossing time (on the order of $1/4$ orbit), and is
dynamically insignificant for the subsequent evolution of the binary,
and in particular the gravitational waves that are generated. 
Here we study three examples from these evolutions --- a 
CP $d=16$ case ($d$ labels the initial separation between
the binaries~\cite{CP}), and two SFCBs,
one equal mass, the other with a mass ratio of $1.5:1$. The
two former evolutions exhibit roughly $2.5$ orbits before merger, the
latter unequal mass case about $1.5$ orbits. In all cases the remnant
is a Kerr black hole with spin parameter $a\approx0.7$, and roughly
$3\%-5\%$ of the rest mass energy of the each system is radiated away
in gravitational waves.  Figure~\ref{f:waveforms} shows a few samples of
the waveforms extracted from these simulations, while
Fig.~\ref{f:cnv_waveforms} demonstrates the convergence behavior of
the wave with resolution for the CP case. In
Fig.~\ref{f:waveforms}, data from the highest resolutions available are
shown.

Observe that the waveforms depicted in Fig.~\ref{f:waveforms} have
some noticeable differences. By construction the phases all match
at $t=0$. On axis, the CP and unequal mass SFCBs
have similar phase and amplitude evolution of the waveform
several cycles to the left and right of $t=0$, however, moving toward the
orbital plane the similarities are less evident. There is also a more
rapid decoherence between the equal mass SFCB
and the other two cases moving away from $t=0$. One reason for
this is that the equal mass SFCB example has initial
conditions tuned to exhibit some ``zoom-whirl'' type behavior, showing
a couple of whirl orbits before the system
finally merges. During the whirl phase the binaries are quite
close together [inside of what might be considered an inner-most
stable circular orbit (ISCO)],
and moving faster than a corresponding point in a quasi-circular
inspiral. Hence the amplitude of this portion of the wave is
quite a bit larger than the quasi-circular case, and remains similar
in magnitude over a couple of wave cycles. 
These differences are in contrast to quasi-circular
equal mass inspiral results obtained by most groups, 
similar ``visual'' comparisons of which suggest remarkable similarity in the
waveforms over several wave cycles away from the 
matching point~\cite{NFNR,NRDA}. Given that {\em all}
the latter evolutions are approximations to {\em essentially the same}
astrophysical scenario, the similarity is not too surprising, but
nevertheless reassuring.

\begin{figure}
\includegraphics[width=3.3in]{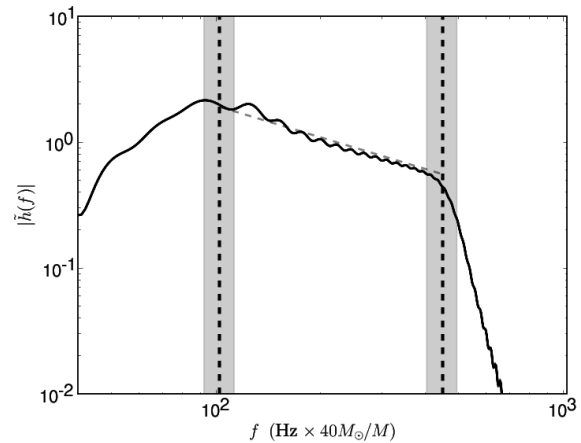}
\caption{\label{f:spectrum} The amplitude of the Fourier transform of
the gravitational waveform, from the evolution of Cook-Pfeiffer
initial data, shown in Fig.~\ref{f:waveforms}.  The vertical dashed
lines are the estimated frequency of the inner-most stable circular
orbit given in Eq.~(\ref{e:fisco}) and the frequency of the dominant
quasi-normal mode, assuming $a=0.7$, given in Eq.~(\ref{e:fqnr}). The
gray shaded region indicates variations in this frequency due to 10\%
changes in the mass used.  Notice how the
power in these waves is predominantly emitted between these two
frequencies; the initial data is such that the binary is orbiting at
or near the ISCO frequency.  In addition, the dashed Gray line which follows the
amplitude is proportional to $f^{-5/6}$. While this is a convincing
fit to the amplitude, we note that there is weak evidence for two
power laws $f^{-7/6}$, as given by post-Newtonian
approximations~\cite{Blanchet:1996pi}, below $\sim 1.5
f_{\mathrm{isco}}$ and $f^{-5/6}$ above that frequency. These
simulations do not cover the inspiral phase well enough to confirm
this result.}
\end{figure}

\begin{figure}
\includegraphics[width=3.3in]{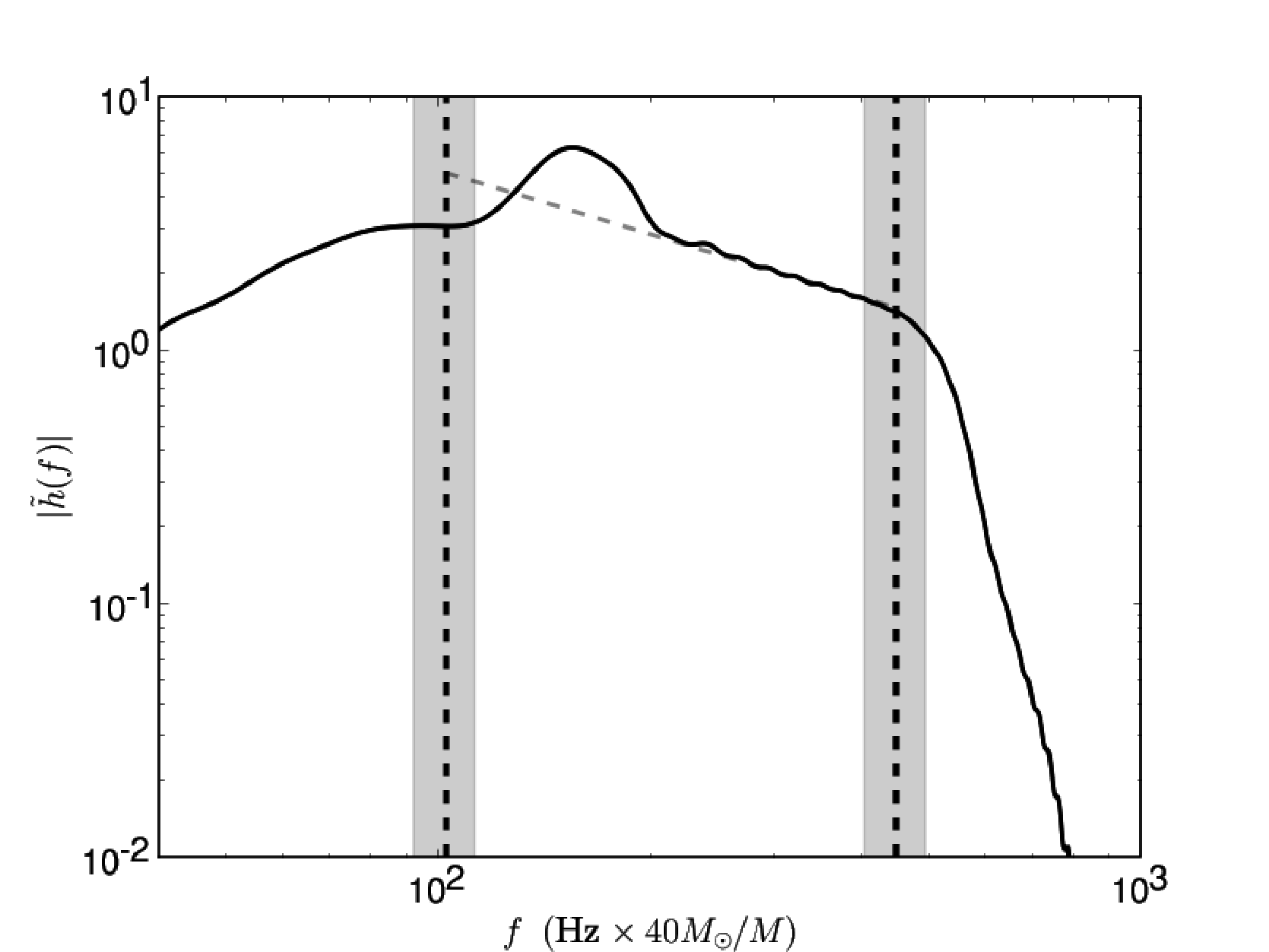}
\includegraphics[width=3.3in]{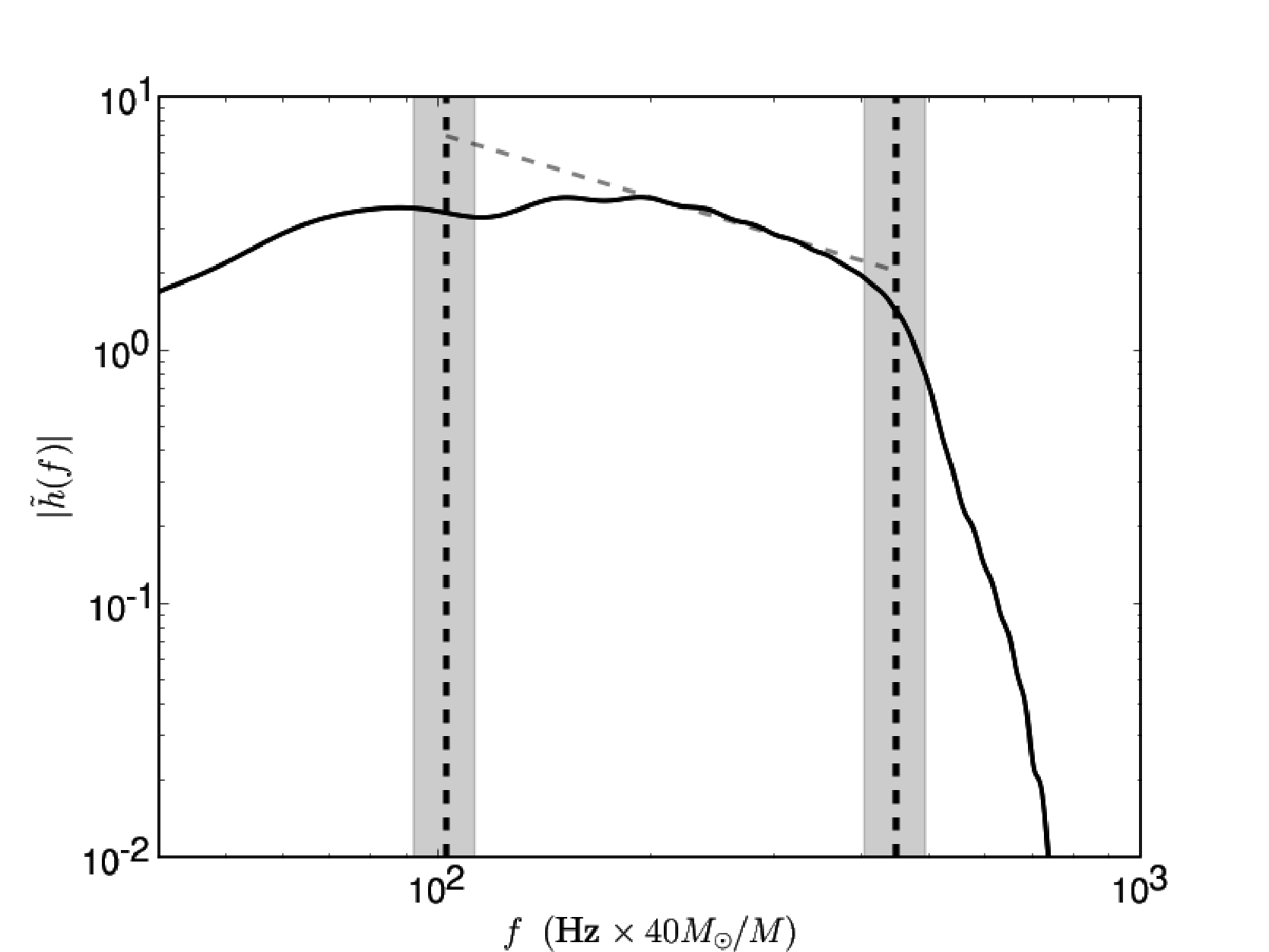}
\caption{\label{f:sfcbspectrum} The amplitude of the Fourier transform
of the gravitational waveforms shown in Fig.~\ref{f:waveforms}: the
top panel is an equal mass SFCB, and the lower
panel is a mass ratio $1.5:1$ SFCB.  Note the
differences between these spectra and that shown in Fig.~\ref{f:spectrum}.
The bump in the equal mass spectrum arises from the hangup of the
binary at roughly constant separation the a brief whirl phase
prior to merger.
While these waveforms are unlikely to be
realized in astrophysical scenarios,  these spectra provide an example
of the rich phenomenology that may be realized in binary mergers.
}
\end{figure}

It is also interesting to examine the Fourier spectrum of these
waveforms and to notice the similarities and differences that are
manifest.  The amplitude of the Fourier transform of
the gravitational waveform, from the evolution of Cook-Pfeiffer
initial data, shown in Fig.~\ref{f:waveforms} is plotted in
Fig.~\ref{f:spectrum}.  This is computed directly from $\Psi_4(t)$
using the frequency domain equivalent of Eq.~(\ref{Psi4_hpc_def})
which gives, for example, 
\begin{equation}
- 4 \pi^2 f^2 \tilde{h}_+(f) = \tilde{\Psi}_4^R(f)
\end{equation}
where $\tilde{\Psi}_4^R(f)$ is the Fourier transform of the real part
of $\Psi_4$. This avoids introducing artifacts from the numerical integration of 
Eq.~(\ref{Psi4_hpc_def}).  To guide the eye, we indicate the
frequency of the inner-most stable circular orbit estimated by Kidder 
et al.~\cite{Kidder:1992fr}
\begin{equation}
\label{e:fisco}
f_{\mathrm{isco}} \approx 205 \left( \frac{20.0 M_\odot}{M} \right) \textrm{
Hz}
\end{equation} 
and the frequency of qausi-normal ringing given by 
\begin{equation}
\label{e:fqnr}
f_{\mathrm{qnr}} \approx 1600 \left[ 1 -
0.63 (1.0 - a)^{0.3} \right] \left( \frac{ 20 M_\odot }{ M_{\mathrm{BH}} } \right)
\textrm{ Hz}
\end{equation}
where $M$ is the total mass of the binary, and $M_{\mathrm{BH}}$ and
$a$ are the mass and spin of the final black hole.  Notice how the
power in these waves is predominantly emitted between these two
frequencies--the initial data is such that the binary is orbiting at
or near the ISCO frequency.  Morevover, the spectrum shows a power-law
spectrum reminiscent of the spectrum of post-Newtonian waveforms.
Similar plots are shown in Figs.~\ref{f:sfcbspectrum} for the SFCB
waveforms.  It is interesting to note the bump in the spectrum when
$m_1=m_2$; this appears to be caused by the zooming orbits referred to
above. It points to the variety of waveform morphologies that might
exist for binary black hole mergers when eccentricity and even spin
of the black holes is large. 

The preceding discussion of similarities and differences between
waveforms is quite heuristic and subjective, and thus of arguable
merit. One of the main purposes of this paper is to propose metrics 
to both quantify the accuracy of simulations and the 
similarities/differences between waveforms, though primarily from the
perspective of data analysis. The sets of waveforms
depicted in Figs.~\ref{f:waveforms} and \ref{f:cnv_waveforms},
that {\em appear} to be significantly different due to either
numerical resolution effects or different initial
physics parameters, will provide useful test cases to gauge
the efficacy of the proposed metrics.

\section{Estimating the waveform accuracy} 
\label{s:accuracy}
 
The accuracy of numerical solutions is generally determined by 
comparing results obtained at different grid resolutions.  This 
approach determines the point-wise convergence of the solutions.  
In the context of gravitational-wave astronomy, only the waveforms 
themselves are directly accessible to observation. It is therefore 
important, if the results of numerical simulations are to be useful to 
gravitational-wave astronomers, to provide a measure of the waveform 
accuracy.   

\subsection{Data analysis formalism}
\label{ss:formalism}

The standard tool of gravitational-wave data analysis is the matched 
filter.  If a signal is present, the detector output is 
\begin{equation} 
s(t) = n(t) + h(t;\hat{\mathbf{\lambda}}) 
\end{equation} 
where $n(t)$ is the noise and $h(t;\hat{\mathbf{\lambda}})$ is the signal. 
In general, the signal depends on a set of unknown (in advance) parameters
$\hat{\mathbf{\lambda}}$.  
We assume the noise is a zero mean $\langle n(t) \rangle = 0$ and
stationary process, i.e. $\langle n(t) \rangle = 0$ and $\langle n(t_1) 
n(t_2) \rangle = Q(|t_1-t_2|)$ for some function $Q$.  
Here $\langle \ldots \rangle$ denotes the ensemble average 
over different instantiations of the noise. 

Define the Fourier 
transform of the noise by 
\begin{equation} 
\tilde{n}(f) = \int_{-\infty}^{\infty} n(t) e^{-2 \pi i f t} dt \; . 
\end{equation}
Since the Fourier
transform is just a linear transformation of the noise time-series, we
also have $\langle \tilde{n}(f) \rangle = 0$. Further, the
variance or power spectrum $S_n(|f|)$ is defined by
\begin{equation} 
\langle \tilde{n}(f) \tilde{n}^\ast (f') \rangle = \frac{1}{2} S_n(|f|) 
\delta(f-f') \; .
\end{equation} 
Notice that this two point function is diagonal;  this is a direct
consequence of the stationarity of the noise. Moreover, the power
spectrum of the noise will depend non-trivially on frequency if the
time-domain correlation function is not diagonal. Noise with a
frequency dependent power spectrum is often referred to as colored
noise. 

In its simplest form,
matched filtering is cross-correlating a template waveform 
$w(t;{\mathbf{\lambda}})$ with the 
time series $s(t)$ observed by the gravitational-wave detector. The
matched filter signal to noise ratio (SNR) is
\begin{equation} 
\rho({\mathbf{\lambda}}) = \frac{2}{\sigma_w}  
\int_{-\infty}^{\infty} df \frac{ \tilde{s}(f) 
\tilde{w}^\ast(f;{\mathbf{\lambda}}) }{ S_n(|f|) } \; , 
\label{e:snr}
\end{equation} 
where  
\begin{equation} 
\sigma^2_w = 2 \int_{-\infty}^{\infty} df \frac{ \tilde{w}(f;{\mathbf{\lambda}}) 
\tilde{w}^\ast(f;{\mathbf{\lambda}}) }{ S_n(|f|) }   
\; . 
\end{equation} 
Notice that the signal-to-noise is normalized such that $\langle 
\rho^2 \rangle = 1$ if the signal is absent, i.e. $h(t; \hat{\lambda})\equiv0$.

Since the signal parameters are not known in advance, one must search
over all possible values of the parameters $\mathbf{\lambda}$ to find
the template that best matches the signal buried in the noise. Then
the output of a search (over a small chunk of data) will be
\begin{equation}
\rho = \max_{\mathbf{\lambda}} \rho(\mathbf{\lambda}) \; .
\end{equation}
The largest SNR will be obtained when 
$w(t;{\mathbf{\lambda}}) = \textrm{constant } \times 
h(t;\hat{\mathbf{\lambda}})$ and $\mathbf{\lambda}=\hat{\mathbf{\lambda}}$.  
In practice, two important issues arise.
First, only a discrete set of values of $\mathbf{\lambda}$ can be
searched. This leads to the notion of a bank of templates with
different parameter values. There is a well developed formalism for
constructing template banks~\cite{Owen:1995tm} for gravitational-wave data
analysis. In Sec.~\ref{ss:comparisons}, we will comment on the number of templates
needed for the binary black hole problem and hence the number of
accurate numerical simulations that must be done. Second, the
theoretical template waveforms may not accurately agree with the real
signals. It is this issue which we address here.

The expected SNR is then 
\begin{equation}
\langle \rho(\lambda) \rangle = \frac{2}{\sigma_w}
\int_{-\infty}^{\infty} df \frac{ \tilde{h}(f;\hat{\mathbf{\lambda}}) 
\tilde{w}^\ast(f;{\mathbf{\lambda}}) }{S(|f|)} \; .
\end{equation}
If the template and the signal are the same, then the optimal SNR is 
\begin{equation}
\langle \rho_{\mathrm{opt}} \rangle = 
\frac{2}{\sigma_h}  
\int_{-\infty}^{\infty} df \frac{ | \tilde{h}(f;\hat{\lambda}) |^2 
 }{ S_n(|f|) } = \sigma_h \; .
\end{equation} 
Define the match~\cite{Owen:1995tm} between a waveform and a template by
\begin{eqnarray}
\mu &=& \frac{\langle \rho(\lambda) \rangle }{ 
\langle \rho_{\mathrm{opt}} \rangle } \\
&=& \frac{2}{\sigma_h \sigma_w}  
\int_{-\infty}^{\infty} df \frac{ \tilde{h}(f;\hat{\mathbf{\lambda}}) 
\tilde{w}^\ast(f;{\mathbf{\lambda}}) }{S(|f|)} \; .
\end{eqnarray}
The \emph{fitting factor}~\cite{Apostolatos:1995pj}
\begin{equation}
\textrm{FF} = \max_\lambda \mu
\end{equation}
is a measure of the distance between a signal and the whole template
space. If $\textrm{FF} = 1$, the signal lies inside the template
space. We will generally use the term match to mean something that is
maximized over a subset of the parameters $\lambda$ and keep fitting
factor for the case when all template parameters have been maximized
over.

\subsection{Re-parametrization of the numerical templates}
\label{ss:reparameterization}

In general, the response of a gravitational-wave detector to the waves
from a compact binary merger will be non-trivial as expressed in
Eq.~(\ref{e:detector-strain}).  Motivated by our intuitive expectation
that gravitational radiation is dominated by the quadrupole waves, we
have explored the following re-parametrization of the numerical waveforms
\begin{eqnarray}
w(t;{\mathbf{\lambda}}) &\approx& A \left( \frac{1 \textrm{ Mpc}}{D}
\right)
\bigl[ \cos\Phi \hat{e}_+( t- t_0; m_1, m_2) \nonumber \\
&& \mbox{\hspace{0.25in}}+ 
\sin\Phi \hat{e}_{\times}( t- t_0; m_1, m_2) \bigr]
\label{e:reparametrized}
\end{eqnarray} 
where $A$ and $\Phi$ depend on the right ascension $\alpha$,
declination $\delta$, polarization $\psi$, inclination $i$ and time
$t$. (The variation in these constants over the short duration of the
signal is completely negligible, however the response of the
instrument to a gravitational wave from a given location on the
celestial sphere depends on the time of day.)  Here the plus and cross
polarization states $\hat{e}_{+,\times}( t- t_0; m_1, m_2)$ are just
the waveforms extracted on the axis orthogonal to the plane of the
binary orbit.  Here we present evidence that this re-parametrization 
may capture the essential features of the merger waveforms
sufficiently well for the purposes of detecting the waves.  On the
other hand,  the full waveforms will probably be needed to extract all
the possible science. 

To see that this approximation is good enough for detection, we re-express 
the matched filtering 
SNR in terms of these templates.  First, notice that the amplitude 
$A \times (1 \textrm{ Mpc} / D)$ cancels out of the SNR defined
in Eq.~(\ref{e:snr}).  Hence the SNR can be expressed as
\begin{equation}
\rho = \max_{\Phi, t_0, m_1, m_2} \frac{z_+ \cos \Phi + z_\times
\sin \Phi }{\overline{\sigma}_w} 
 \; ,
\end{equation}
where $\overline{\sigma}_w = D \sigma_w / ( A \times 1 \textrm{ Mpc} )$ and
\begin{equation}
z_{+,\times}(t_0; m_1, m_2) = 2 \int_{-\infty}^{\infty} \! \frac{ \tilde{s}(f) 
\tilde{e}^\ast_{+,\times}(f;m_1, m_2)  }{ S_n(|f|) } e^{2 \pi i f t_0}
df
\; . 
\end{equation}
For the waveforms considered here, we find that 
\begin{equation}
2 \int_{-\infty}^{\infty} df \frac{ |\tilde{e}_+(f;m_1, m_2)|^2
}{ S_n(|f|) } 
 \approx 
2 \int_{-\infty}^{\infty} df \frac{ |\tilde{e}_{\times}(f;m_1, m_2)|^2
}{ S_n(|f|) } 
\end{equation}
to better than 3\% accuracy. We also find that the two polarizations
are almost orthogonal, that is  
\begin{equation}
\int_{-\infty}^{\infty} df \frac{ \tilde{e}_+(f;m_1, m_2)
\tilde{e}^\ast_{\times}(f;m_1,m_2) }{ S_n(|f|) } \approx 0 \; ,
\end{equation}
with typical values $\sim 3 \times 10^{-3}$.  Hence the
normalization constant $\sigma_w$ simplifies considerably and is
independent of $\Phi$ and $t_0$:
\begin{equation}
\sigma^2_w \approx 2 \int_{-\infty}^{\infty} df \frac{ |\tilde{e}_+(f;m_1, m_2)|^2
}{ S_n(|f|) } \;  .
\end{equation}
This allows us to maximize over $\Phi$ analytically to find
\begin{equation}
\rho = \max_{t_0, m_1, m_2} \sigma_w^{-1} \sqrt{ z_+^2(t_0; m_1, m_2) + 
z_\times^2(t_0; m_1, m_2) }  \; .
\label{e:quadrature-snr}
\end{equation}
Below, we will use the plus and cross quadrature matched filters to
quantify both the accuracy of the numerical simulations and the
accuracy of the approximation introduced in Eq.~(\ref{e:reparametrized}).

Moreover, the match can be written as
\begin{equation}
\mu = \max_{\Phi, t_0, m_1, m_2} ( \mu_+ \cos\Phi + \mu_\times \sin\Phi )
\label{e:match}
\end{equation}
where 
\begin{equation}
\mu_{+,\times} = \frac{\langle z_{+,\times} \rangle}{\sigma_h
\sigma_e} \; . 
\label{e:matchplustimes}
\end{equation}

It is straightforward to see that the validity of the re-parametrization in 
Eq.~(\ref{e:reparametrized}) requires that $\mu_+ \approx 1$ and $\mu_\times
\approx 0$, independent of the inclination $i$, when $h(t,t_0,m_1,m_2) = 
h_+(t, t_0, m_1, m_2, i)$, and that $\mu_\times \approx 1$ and $\mu_+
\approx 0$, independent of the inclination $i$, when $h(t,t_0,m_1,m_2) =
h_\times(t, t_0, m_1, m_2, i)$.  This result is
confirmed in Table~\ref{t:match-v-angle}; only the mass space needs to be
searched by the explicit construction of a discrete bank of templates.

It must be stressed, however, that this approach is not sufficient for
the interpretation of observations and measurement of parameters.
Moreover, there are likely to be regimes where this approach is
insufficient even for detection.

\begin{table}
\caption{\label{t:match-v-angle}
The range of matches between waveforms extracted at different angles
relative to the binary and a template given by the waveform extracted
at the axis for masses $40 M_\odot < M < 200 M_\odot$.  Note that the 
$++$ and $\times\times$ entries are all
very close to unity. The $+\times$ entries are close to zero
indicating that the $+$ and $\times$ polarizations are almost
orthogonal.  This suggests that the reparametrization in
Eq.~(\ref{e:reparametrized}) could be good enough for detection purposes, with
LIGO, in the equal mass binary case.}
\begin{tabular}{l|c|c|c|c}
{CP} & \multicolumn{2}{c}{$h_+(t_0, m_1,m_2,i)$} &
\multicolumn{2}{c}{$h_\times(t_0, m_1,m_2,i)$}\\
\hline
$i$  & $3 \pi / 8$ & $\pi / 4$ & $3 \pi / 8$ & $\pi / 4$ \\
\hline
$\mu_+$      & [0.980,0.995] & [0.990,0.996] & [0.017,0.044]  & [0.044,0.066]\\
$\mu_\times$ & [0.050,0.074] & [0.046,0.069] & [0.989,0.995] & [0.992,0.996] \\
\hline
\hline
{SFCB} & \multicolumn{2}{c}{$h^{\mathrm{SFCB}}_+(t_0, m_1=m_2,i)$} &
\multicolumn{2}{c}{$h^{\mathrm{SFCB}}_\times(t_0, m_1=m_2,i)$}\\
\hline
$i$  & $3 \pi / 8$ & $\pi / 4$ & $3 \pi / 8$ & $\pi / 4$ \\
\hline
$\mu_+$      & [0.981,0.989] & [0.993,0.996] & [0.021,0.044]  & [0.019,0.032]\\
$\mu_\times$ & [0.051,0.075] & [0.024,0.038] & [0.995,0.989]  & [0.995,0.998]\\
\hline
\hline
{SFCB} & \multicolumn{2}{c}{$h^{\mathrm{SFCB}}_+(t_0, m_1 = 1.5 m_2,i)$} &
\multicolumn{2}{c}{$h^{\mathrm{SFCB}}_\times(t_0, m_1 = 1.5 m_2,i)$}\\
\hline
$i$  & $3 \pi / 8$ & $\pi / 4$ & $3 \pi / 8$ & $\pi / 4$ \\
\hline
$\mu_+$      & [0.969,0.973] & [0.986,0.991] & [0.042,0.056]  & [0.040,0.050]\\
$\mu_\times$ & [0.076,0.096] & [0.049,0.061] & [0.987,0.977]  & [0.989,0.993]\\
\end{tabular}
\end{table}

\subsection{Example of waveform accuracy estimation}
\label{ss:example}

As we mentioned above, differences between the true waveform from a
binary black hole merger and the template waveform can result in
degradation of the SNR.  In particular, numerically generated waveform
templates might be inaccurate for any number of reasons, e.g.
truncation errors, instabilities, errors in boundary conditions, or
incorrect waveform extraction. To get a handle on these effects, it is
useful to compare waveforms which are supposed to represent the same
physical process that were generated in different ways.

\begin{figure}
\includegraphics[width=3.3in]{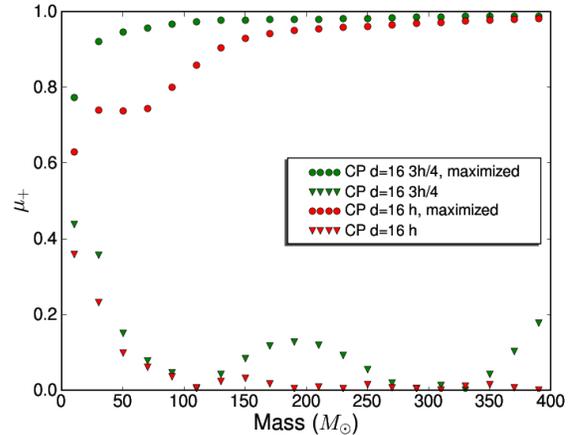}
\caption{The match between the $h_+$ waveform polarization computed as
the overlap between the waveforms extracted from two different
resolution runs. Circles indicate the match when the waveforms are
allowed to shift in time relative to each other.  The triangles
indicate the match computed as the overlap integral between waveforms
without the time shift.  The highest resolution simulation is used as
the reference and the match with two coarser resolution simulations
are computed.}
\label{fig:match-v-mass}
\end{figure}

Here, we present a sample analysis of the waveforms presented 
in~\cite{Pretorius:2005gq,pretoriusmomentum,buonanno_et_al}. Using the
formalism outlined above, one can compute the matches $\mu_+$ and
$\mu_\times$ for the waveform generated at the finest resolution and
templates at coarser resolutions.  This investigation follows the
standard convergence testing of numerical relativity, but with an
emphasis on the utility of the waveforms for data analysis.  Moreover,
the answer to this question depends on the mass of the binary and
the detector being considered (as it depends on its particular noise
curve).  In Fig.~\ref{fig:match-v-mass}, we show the
match $\mu_+$ versus the binary mass for the initial LIGO noise curve shown in
Fig.~\ref{f:ligo-noise}.
For each template waveform (i.e. the
waveforms from evolutions with coarser resolution), there are sets of points: the
triangles indicate the match before maximization over $t_0$, the
circles indicate the match after maximization over $t_0$.  When the
match exceeds $0.9$,  we may be tempted to conclude that the finest
resolution waveforms are sufficiently accurate to be used for
gravitational-wave data analysis.  This is not the whole story,
however.  Referring back to Fig.~\ref{f:spectrum}, we note that
the waves with frequencies $f \alt 205 ( 20 M_\odot / M)$ appear to
depend on the initial data while those above that frequency might
reasonably be considered independent of the initial data.   
Given that the current LIGO instruments
are sensitive to waves above $40 \textrm{ Hz}$, this suggests that 
the match should only be trusted for masses $M \agt 100 M_\odot$.

\begin{figure}
\includegraphics[width=3.3in]{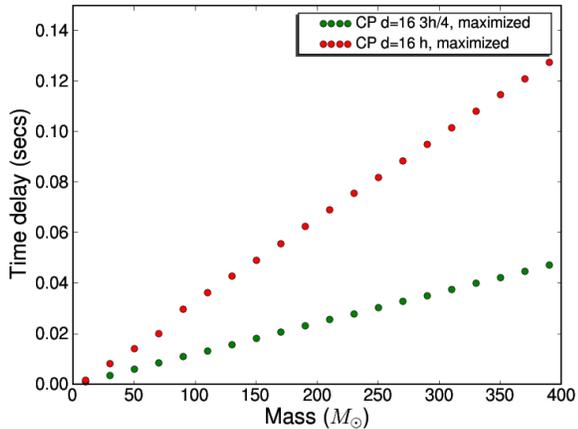}
\caption{The time-offset which maximizes the match (for the same mass) 
between the $h_+$ waveform polarization computed as
the overlap between the waveforms extracted from two different
resolution runs. The green uses the finest simulation as the waveform
and the coarsest simulation as the template; the red uses the finest 
simulation as the waveform and the next finest simulation as the template.
While this shows convergence, there may still be systematic offsets
between these waveforms and real physical waveforms.}
\label{fig:time-delay-v-mass}
\end{figure}

On the other hand, the small match obtained without maximization over
time suggests that the evolutions are slightly different.  In
Fig.~\ref{fig:time-delay-v-mass}, we show the time shift needed to
maximize the match.  Note that the shift scales linearly with mass.
This suggests that the difference between these waveforms might be
captured simply by rescaling the mass of the template, i.e. maximizing
over both total mass and time-delay as one would do in a search. In a
simple simulation which searched over various template masses for a
fixed waveform mass, it was found that the best match is achieved when
the template mass is different from the waveform mass,  but the
time-shift is similar in magnitude to that obtained when using the
same mass in both the template and the waveform.

In this concrete example, we have compared the waveforms from the same
simulation at different resolutions. While the waveforms generated at
the finest resolution appear to be accurate enough to use as templates
in searches for gravitational waves, the systematic differences
between the mass and $t_0$ which maximizes the match hint that the
waveforms may not be faithful to the physical system.  That is, the
map between mass of the template and mass of the binary (in the
standard Newtonian sense) may have systematic biases.  It will be
important to explore these issues by comparing waveforms from
simulations (starting from the same initial data) by different groups.
Moreover, detailed exploration of the dependence of the waveforms on 
the initial data will also bring information about the faithfulness of
the waveforms\footnote{Related metrics have been discussed recently to estimate
the needed accuracy of the simulated waveforms and remove numerical errors
as well as computational grid effects\cite{markmiller}.}.

\subsection{Comparing waveforms from different simulations}
\label{ss:comparisons}

In the previous section, we presented a comparison between waveforms
extracted from simulations at different resolutions in order to gain
insight into the accuracy of the numerical solutions.  Another
important step in exploring the full parameter space of compact binary
inspiral for earth-based detectors is to compare the results of
different simulations.  There are two different reasons for making
these comparisons:  First, comparison of waveforms representing the 
same physical solution carried out using different methods will allow
a deeper understanding of the numerical issues in this very
complicated simulation problem.  Second, it may allow more efficient
exploration of the parameter space if multiple groups can agree on
some key test cases and then explore, in detail, other regions of
parameter space. 

\begin{figure}
\includegraphics[width=3.3in]{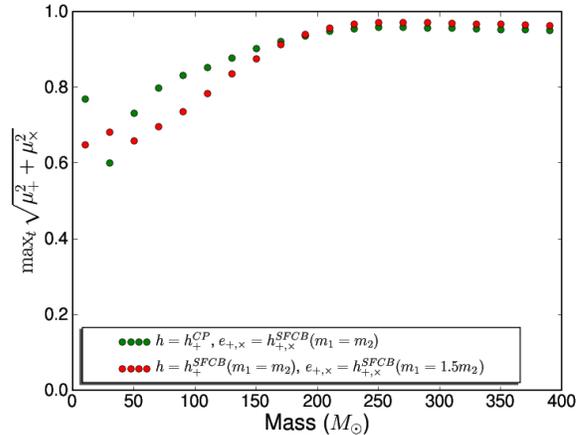}
\caption{The match between the $h_+$ waveform polarization from one
simulation with the waveforms from another simulation maximized over
both time and phase. The match $\max_t\sqrt{ \mu^2_+ + \mu^2_\times}
\agt 0.95$ for masses $M \agt 200 M_\odot$.  This gives a quantitative
measure of the similarities and differences between the waveforms
presented in Figs.~\ref{f:waveforms}--\ref{f:sfcbspectrum}.
}
\label{f:inter-simulation}
\end{figure}

As an example, we compare the waveforms from the CP simulations with
those from the SFCB simulations and between the different SFCB
simulations in Fig~\ref{f:inter-simulation}.  The results quantify the
degree to
which these waveforms are different/similar as seen in
Figs.~\ref{f:waveforms}--\ref{f:sfcbspectrum}.  For example, the match
(maximized over time and phase) between $h^{\mathrm{CP}}_+$ and the
equal-mass SFCB waveforms is greater than $0.95$ for masses $M \agt
200.0$.  This is consistent with agreement in both frequency spectrum
[$f \agt 200 (40 M_\odot/M)$] and the waveform. The biggest difference
between these waveforms therefore appears to come from the
eccentricity of the binary orbit when the black holes first form in
SFCBs. A similar conclusion holds for the match (maximized over time and phase) 
between $h^{\mathrm{SFCB}}_+(m_1=m_2)$ and the SFCB waveform for mass
ratio $m_1/m_2 = 1.5$.

Finally, we note that the number of templates needed to \emph{search}
for gravitational waves from \emph{non-spinning binaries in data from
earth-based detectors} can be estimated as follows.  From the CP
simulations, we find that the match (maximized over time and phase) is
$\mu \geq 0.97$ for two waveforms with masses differing by $\approx
0.05 M$ with $100 M_\odot < M < 400 M_\odot$.  Templates along the
equal mass line would then be laid out with separations $\approx 0.1
M$ giving $\approx \int_{100}^{400} dM/ (0.1 M) \approx 13$ templates.  
If we construct a square grid on the $m_1m_2$-space by drawing lines through
the equal-mass template points, this gives $\approx 91$ templates
to cover the square defined by $50 M_\odot < m_1, m_2 < 200 M_\odot$.
This formula assumes a maximum reduction in SNR of $3\%$ due to
template mismatch.  Since a single numerical simulation gives all
templates along a line of constant mass ratio, this suggests that
about 10 simulation runs would be needed to cover this mass space
densely enough for detection purposes. This crude
estimate should be refined with more detailed simulations with
different mass ratios and larger initial separations. It is important
to stress here that this number assumes that all the physics and numerical
issues are under complete control, that no tuning or test runs need to be
made, and that the waveforms do not get much more complicated as the
mass ratio changes. All of these issues need to be explored before one
could generate waveform templates that could be confidently used in
gravitational wave detection. As a result, while the final number
of templates is not very large, reaching the stage where these
runs can be made requires much greater effort. 

It is important not to read too much into the estimate stated above.
As we have emphasized, it ignores spin and many other important issues
in numerical relativity.  It is also only applicable to earth-based
detectors. The problem is different for higher mass binaries in the
LISA band.  Finally, it only estimates the number of simulations
needed to enhance detection.  As we discuss later, this is the first
step in using numerical methods to extract scientific information from
gravitational-wave observations, but the larger computational task
will be extracting accurate information from the data once a detection
is made.  

\subsection{Detectability of numerical templates using ringdown filters}
\label{ss:ringdown}

It is illustrative to determine how well existing methods of searching for
gravitational wave bursts would work in detecting numerical waveforms.
In particular, we would expect that ringdown waveform matched filters would
work well at detecting the end of the numerical waveforms (which do
correspond to black hole quasi-normal mode ringdown)---especially if it is
this portion of the waveform that is in the detector band.  However, for
a broad range of masses it is the merger waveform that is in LIGO's sensitive
band rather than the ringdown waveform: indeed, for the most likely ranges
of binary black hole masses, the ringdown radiation will be at frequencies
higher than the most sensitive portion of LIGO's band.  Nevertheless, the numerical
waveforms might be well-matched by a ringdown template even
at frequencies below those of the final black hole's quasi-normal mode.
Will a ringdown matched filter template actually do well at detecting these
numerical waveforms?
Also, will the presence of a preceding waveform bias
ringdown extraction parameters?  The answer to both of these questions seems to
be yes.

The match, $\mu$, can be used to measure the ability of the ringdown filter
to detect the waveform.  In the case of a ringdown filter the relevant
parameters that we must maximize in forming the match are the start time
of the ringdown, $t_0$, the central frequency of the ringdown, $f_0$, and
the ringdown quality factor $Q$ which measures the decay time of the ringdown
in cycles.
A ringdown waveform is an
exponentially-damped sinusoid: $\exp(-\pi f_0 \tau/Q)\cos(2\pi f_0\tau)$
for $\tau=t-t_0>0$.  As before, the match is an indication of the fraction
of the signal-to-noise ratio that a ringdown filter will obtain compared
to an optimal filter.  The match will depend on the mass of the waveform
as this determines which portion of the waveform is in LIGO's band.
From Fig.~\ref{f:spectrum} it can be seen that the numerical waveforms do
not accurately give the gravitational wave energy below 40~Hz for a 
$100\,M_\odot$ black hole.  Since 40~Hz is roughly the low-frequency bound 
of LIGO's sensitive band, care must be taken in interpreting the match when
using low mass waveforms with a total mass less than $\approx100\,M_\odot$
since a substantial contribution to the gravitational wave signal is missing in
LIGO's sensitive band.

\begin{figure}
\includegraphics[width=\linewidth]{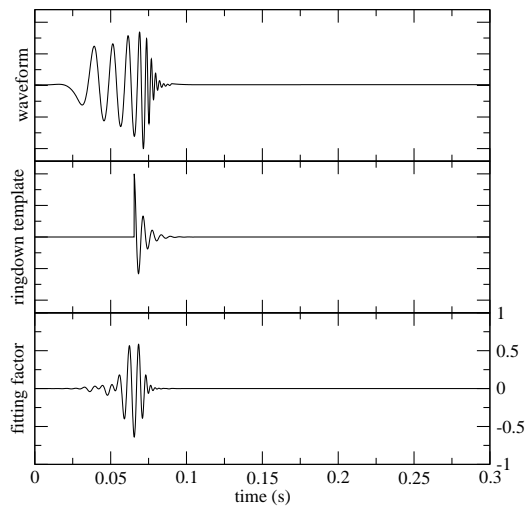}
\caption{The $h_+$ numerical waveform for a total mass of $50\,M_\odot$
(top), the best-matching ringdown (exponentially-damped sinusoid) filter
(middle), and the result of filtering the waveform with this filter (bottom),
all as functions of time.
The match, which is the largest absolute value of the bottom plot, is 64\%.
The best match clearly occurs before what we would call the ringdown phase of
the numerical simulation: this is because the ringdown phase is not in LIGO's
sensitive band for a total mass of $50\,M_\odot$---the ringdown filter
therefore obtains its best match at an earlier (and lower frequency) portion
of the numerical waveform.   Although the ringdown filter does not do too
badly at detecting the numerical waveform, it could not be used to measure
the quasi-normal mode frequency of the final black hole without using
information from both the ringdown and merger phases.  This serves to
emphasize the importance of developing data analysis techniques which
combine information from all three phases of binary
evolution~\cite{saikat-thesis}.}
\label{fig:ring-50-msun}
\end{figure}

\begin{figure}
\includegraphics[width=\linewidth]{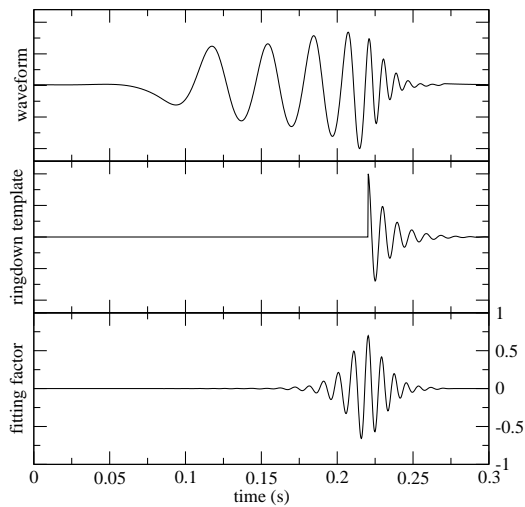}
\caption{The $h_+$ numerical waveform for a total mass of $150\,M_\odot$
(top), the best-matching ringdown (exponentially-damped sinusoid) filter
(middle), and the result of filtering the waveform with this filter (bottom),
all as functions of time.
The match, which is the largest absolute value of the bottom plot, is 70\%.}
\label{fig:ring-150-msun}
\end{figure}

Figures~\ref{fig:ring-50-msun} and \ref{fig:ring-150-msun} illustrate the ringdown filter response to the numerical
waveforms for two masses, $50\,M_\odot$ and $150\,M_\odot$.  In these figures
there are three panels which show (top) the numerical strain waveform;
(middle) the best-matching ringdown template; and (bottom) the result of
filtering the waveform with the best-matching ringdown template---the match
is the maximum absolute value of this trace.  In both cases the match is
substantial: 64\% for the $50\,M_\odot$ case and 70\% for the
$150\,M_\odot$ case.  In the $150\,M_\odot$ case (Fig.~\ref{fig:ring-150-msun}) it is clear that
the best-matching ringdown filter is in fact matching the ringdown portion
of the numerical waveform; however, in the $50\,M_\odot$ case (Fig.~\ref{fig:ring-50-msun}), the
best matching ringdown waveform is matching the numerical waveform considerably
earlier than the ringdown phase.  For the $50\,M_\odot$ case, the ringdown
phase is not in LIGO's sensitive band.  The ringdown filter, while it
performed reasonably well at detecting the earlier phase of the numerical
waveform, could not be used to measure the quasi-normal mode frequency of the
resulting black hole.  Also note that in the $50\,M_\odot$ case, the match
is an overestimate because the numerical waveform is missing a portion of the
late inspiral that would be in LIGO's band for this mass, as remarked above.\\

\section{Discussion and conclusions}
\label{s:discussion}

Recent computational and algorithmic advances in numerical relativity
allow the exploration of strong field general relativity in regimes
which were hitherto inaccessible.  In this paper, we have
presented a measure of the accuracy of these simulations 
adapted to the use of these results in observational
gravitational-wave astronomy. The formalism presented in
Sec.~\ref{ss:formalism} is just the match, defined in Ref.~\cite{Owen:1995tm},  applied
to waveforms from various numerical simulations.  By applying this
metric to the equal-mass Cook-Pfeiffer binary black hole simulations
presented in~\cite{buonanno_et_al}, we conclude that these simulations show
convergence within this measure.  Nevertheless, we also sound a
cautionary note about the match as used in Sec.~\ref{ss:example} to measure
the convergence:  it only measures the convergence, not the physical
relevance of the ultimate solution.  To determine the latter, one must
also examine a host of other issues including the nature of the
initial data, the method of waveform extraction, and the
commonly examined issues of stability, convergence and independence 
on boundary effects. Additionally, the currently available
waveforms cover just a fraction of the relevant sources and analysis
similar to those presented here will need to be carried out as other
cases are treated.

In the remainder of this section we conclude by discussing 
several outstanding issues in the use numerical relativity as a tool for
gravitational-wave astronomy, including faithful extraction of the
waveform from the simulations, what could be the most useful
information that numerical simulations of compact object
interactions could provide in the near-term to enhance the
detectability of gravitational waves, and what information
from the simulations could help
us learn the most about compact objects after detection. 

\subsection{Issues in numerical simulations}

Numerical simulations carried out by different codes by construction,
necessity and available computational resources adopt different formulations,
employ distinct coordinate systems and varied discretization schemes. As a result,
care is required not only in translating the results to data analysis, but also 
comparing results from different codes. Naturally, concentrating on physical observables,
in our particular case gravitational waves, is the sensible way to do this. At the
analytical level, a well defined approach to compute the radiative properties
of the spacetime (under natural assumptions) was developed in the 60's by 
taking advantage of future null infinity (${\scri}^+$)\cite{bondi,sachs,NP}. However,
numerical applications dealing with black hole spacetimes can not yet reach
${\scri}^+$ (this awaits Cauchy-characteristic matching or the Conformal equations
to be fully implemented in the type of systems being discussed here) 
though there is a strong need to compute the radiation
produced in the problem. To do so within a numerical simulation, two approaches are routinely
taken. One is based on perturbative methods\cite{zerilli,regge_wheeler,moncrief,Abrahams:1995gn} 
which relies on
a suitable identification of a background spacetime in particular coordinates and
extracting specific quantities from the simulation. A second approach, which has
become the most common one, makes use  of the
infrastructure developed to calculate the radiation at ${\scri}^+$ but applied at
 a finite distance from the source (see for instance
\cite{Teukolsky:1973ha,smarr,Baker:2001sf,Campanelli:2005ia,qk1,qk2,qk3,bruegmann_et_al} ).  
While in principle this approach can be used beyond the perturbative level
and is less sensitive to identifying a correct background, 
quantities defined at ${\scri}^+$ need to be translated
to finite distances where they may not be well or unambiguously defined. 

Several key elements, listed below, are in general required for faithful
extraction of gravitational waves a finite distance from the source
using the standard result for the relationship between
$\Psi_4$ and the gravitational wave strain (\ref{Psi4_hpc_def})
(we focus here on items pertaining to ${\scri}^+$-based extraction
tools, though similar comments apply to perturbative-based methods).
For the most part in present simulations it is {\em assumed} that these
conditions are satisfied with systematic error less than
numerical truncation error---of course, these assumptions will
eventually need to be verified, and we discuss some suggestions
on how to do this in Appendix \ref{appwaves}. The items listed
below are not all independent. Furthermore, in theory several
of the items are {\em not} strictly required {\em if}
during wave extraction artifacts induced by ``bad'' coordinates
are identified and removed; additional discussion of this is
also presented in Appendix \ref{appwaves}.

\begin{itemize}
\item In the extraction zone the wave travels with unit 
      coordinate velocity, and the amplitude decays as $1/r$.
      If these conditions are not satisfied
      the extracted waves could suffer an error in amplitude 
      and a shift in the frequency of the wave (signs of this
      kind of gauge artifact are seen in the CP evolutions 
      with generalized harmonic gauge\cite{buonanno_et_al}).
\item The extraction world tube is assumed to be a geometric 
      sphere, and more-over it is assumed that the metric of this sphere
      can be expressed as $ds^2 = r^2 [d\theta^2 + \sin^2\theta d\phi^2]$, 
      where $r$ is the extraction radius and $(\theta,\phi)$ are
      the usual spherical polar coordinates
      mapped onto the extraction sphere. Deviations 
      from these assumption could, for example, lead to artificial
      mixing of the spherical harmonic components of the 
      waveform; and of course the correct identification of 
      these harmonic components is an important tool in understanding
      and quantifying the physics of different merger scenarios.
\item Each point on the extraction world tube is assumed to 
      correspond to an inertial observer. Together with the above 
      items this is equivalent to assuming that 
      the lapse function $\alpha=1 + O(1/r)$ and 
      the shift vector induced at the worldtube $\beta^A =0 + O(1/r)$. If these conditions
      are not satisfied all the problems mentioned in the preceding
      items could manifest.
\item Even with a perfect gauge the $O(1/r)$ approach
      to Minkowski space could induce spurious effects at finite
      extraction distance; thus the extraction radius must be far enough 
      from the source that the $O(1/r)$ systematic errors in  
      waveforms are smaller than the numerical truncation error. 
      This issue can be alleviated in part by a judicial choice of
      the tetrad used to calculate $\Psi_4$ \cite{qk1,qk2,qk3}.
\end{itemize}

\subsection{Enhancing the detectability of gravitational waves}

Numerical relativity has long been touted as necessary to doing the
best science with ground based gravitational-wave detectors since the
strongest sources of gravitational radiation involve strong
gravitational fields and the full non-linearity of general
relativity.  In \cite{Flanagan:1997kp},  Flanagan and Hughes laid out the issues
relating to the detection and measurement of waves from binary black
holes.  They conclude that binary black holes in the mass range $25
M_\odot \alt M \alt 700 M_\odot$ may be the strongest sources of
gravitational waves accessible to earth-based detectors.  In this mass
range, they speculated that most of the detectable gravitational-wave energy
would come from the merger waves emitted between $f_{\mathrm{isco}}$
and $f_{\mathrm{qnr}}$ and guessed that about $3\%$ of the binary
mass would be emitted as gravitational waves from the ringdown of the
final black hole.  Numerical relativity simulations can now 
provide a wealth of information about the merger and ringdown phases 
of compact binaries, even though precise connection to the inspiral
phase may remain elusive for some time. 

\begin{figure}
\includegraphics[width=3.3in]{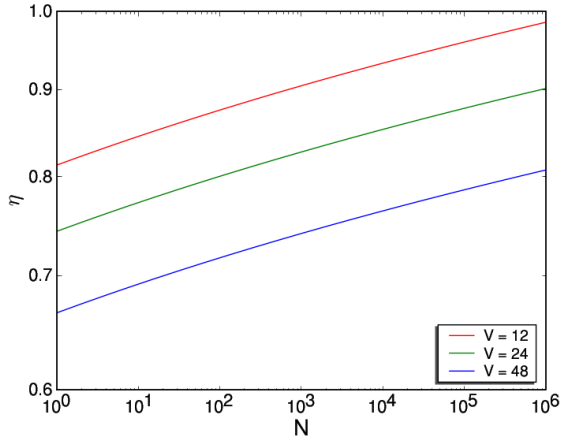}
\caption{\label{f:effectiveness}
The relative effectiveness $\eta$ of an excess-power search with known
time-frequency volume $V$ compared to a matched filtering search
with $N$ effectively independent templates assuming both searches are
tuned to give a $1\%$ false-alarm probability.  The relative effectiveness is defined as 
$\eta = D^{\mathrm{EP}}_{99\%} / D^{\mathrm{MF}}_{99\%}$ where
$D_{99\%}$ is the effective distance at which $99\%$ of the sources are
detectable in each search.  Given the estimated time-frequency volume
of the merger signal from equal mass binaries, i.e. $V\simeq 12$, this 
suggests that a sub-optimal search
strategy could still detect about half the sources that could be
detected in a matched filtering search; simple enhancements to the
excess power method should be able to do better. This data is taken from
Fig.~4 of Ref.~\cite{Anderson:2000yy}. 
}
\end{figure}

With the cautionary note that so far numerical simulations have provided
input on a (small) subset of the physical parameter space and that,
in particular, spin-orbit interactions might strongly influence
the modeled waveforms, valuable insights can be drawn with the
current knowledge.
Consider the simulations of CP initial data 
discussed here.  We can immediately
make several qualitative observations about the waves (see also
the relevant discussions in \cite{buonanno_et_al}).  First, the
waves sweep smoothly upward in frequency from $f_{\mathrm{isco}}$
to $f_{\mathrm{qnr}}$; during this phase the time domain amplitude
also increases monotonically.  About $3$--$5\%$ of the binary mass is
radiated in the last plunge-orbit, merger and ringdown.
The analysis in Sec.~\ref{ss:ringdown}
suggest that upwards of $\sim 70\%$ of the wave energy 
can be attributed to the ringdown waves (though some portion
of the wave matched by the ringdown template could 
be associated with the late-inspiral part of the wave).
Perhaps most important
for detection of the waves is 
that the bulk of the wave energy is emitted in a time-frequency volume
defined by $102.5 (40 M_\odot / M) \alt f \alt 500 (40 M_\odot / M)$
within a time duration $\Delta t \approx 0.03 (M / 40 M_\odot)$.
In the language of Anderson et al.~\cite{Anderson:2000yy},  the time-frequency
volume of the signals is $V=12$ with a definite mass-scaling for the
frequency band. Figure~\ref{f:effectiveness} compares the sensitivity
of a matched filtering search to an excess power search for several
different time-frequency volumes.  In the context of these equal mass,
non-spinning binaries, it suggests that a naive excess-power search
would miss about half of the signals detectable by a matched filtering search
for the merger only.   Nevertheless, this information, when carefully combined with a
search for gravitational waves from the inspiral and ringdown phases of
binary evolution could provide a near optimal search for these
sources. 

Dynamical simulations of binary black holes have so far only been
carried out for a very limited number of parameters (e.g.~mass ratio
and black hole spin), and it will be very important to systematically
investigate the dependence of the waves on these parameters
(compare~\cite{loustospin,pretoriusmomentum}). For example, black hole
spin may effect both the amplitude and the duration of the merger
waves.  For aligned black hole spins the merger takes longer than for
anti-aligned black hole spins \cite{loustospin}.  In addition, the
interaction between spin and orbital angular momentum may play a role
in the dynamics of the merger (even though the speculations of
\cite{PriceWhelan01} have not been confirmed by the dynamical
simulations of \cite{Campanelli_etal_06b}).

An important next step in confirming these speculations is to use the
waveforms from numerical simulations as sample signals in real
detector noise and passed through the current detection pipelines.
This would be facilitated by the development of an archive of
gravitational waveforms in a uniform format that could be used openly
by the gravitational-wave detection community to calibrate their
searches and their pipelines. Activities in this direction are
already under way~\cite{NRwaves};  with the intention of
developing  into a useful resource for
gravitational-wave astronomy.  As an aside, to expedite
their use in data analysis, these waveforms should be
provided as equally sampled time series,  with time in units of
seconds and scaled to a physical distance of $1 \textrm{ Mpc}$ from
the binary.  

Despite recent computational and algorithmic advances, numerical
simulations are costly and will be for years to come.  In Appendix
\ref{app_cost}, we present order of magnitude estimates of the
computational cost of simulations needed to produce parameter space
surveys of a given accuracy and physical evolution time. This suggests
that it will be impossible to populate a template bank solely with the
results of numerical simulations in the near to medium term. It is
therefore interesting to devise approximate methods which might
capture the essential features of the merger phase.  One possible
approach would be to use PN methods for the early phase and close
limit approximations for the late phase with a judiciously chosen
behavior in between.  Adopting such approximate methods requires
a careful understanding of the sensitivity of detection methods
to differences in the approximate waveforms.

\subsection{Learning about compact objects through 
gravitational-wave observations}

While the direct observation of gravitational waves will be a huge
achievement,  we hope that the first detection will only mark the
beginning of gravitational-wave astronomy.  This field is built on the
premise that we can decode information about the sources of
gravitational waves from the signals observed at a detector. To
achieve this goal, we need the ability to simulate the generation of
gravitational waves by various sources. Moreover, it is the imprint
of the sources on the waves that will carry some of the most
interesting information.

\subsubsection{Binary Black Holes}
By exploring the results of merger simulations from the perspective
of data analysis we can obtain a better picture of what we would
be able to learn about compact object interactions from future observations.
Therefore it will be very important in the near future to perform
surveys of a wide variety of initial data parameters (in particular
varying mass ratios and spin vectors), not so much to build
template libraries, but to understand what the broad features
of merger look like through the lense of a gravitational wave detector.
For example, it has long been anticipated that the onset of the binary
merger, at which the slow and adiabatic binary inspiral changes into a
dynamical plunge, would occur at an inner-most stable circular orbit
and would leave a characteristic signature in the gravitational 
wave signal, which could be measured in gravitational-wave detectors.
While the recent dynamical simulations of binary black holes do not
reveal any abrupt change in the waveform, there does appear to be a
break in the frequency spectrum of the waves which occurs somewhere
between the predicted ISCO frequency and the quasi-normal mode
frequency---see Fig.~\ref{f:spectrum}, and further
discussion of this in \cite{buonanno_et_al}. It is certainly plausible
that this break in frequency becomes more pronounced as one moves
away from the equal mass, non-spinning regime.

Black hole coalescence has also been regarded
as giving rise to an arena where the strong-field, non-linear regime of 
general relativity will be clearly revealed to observers. The
early simulations show perhaps a disappointing lack of such features,
where except for a very short and smooth transition between inspiral
and ringdown, much of the waveform can be understood using perturbative
techniques. Another way of stating this is that all waveforms
to date are dominated by the quadrupole harmonic. In general one would
expect non-linear effects to result in mode coupling. Again, that
we do not see significant higher order harmonics could 
be due to the restricted initial conditions so far considered; however, at the
very least this is telling us that manifest strong-field effects
are not ubiquitous in this type of black hole collisions, and the community will
need to search harder to find richer regions
of astrophysically relevant parameter space.
Whether this observation remains as such in more generic cases will have
strong consequences for the simulation and analysis sides. On one hand,
deciphering the non-linear effects would require significantly more accurate simulations
and a considerably denser template bank for data analysis. On the other hand however,
these templates could be parametrized in a
rather simple form like that in Eq.~(\ref{e:reparametrized}).

\subsubsection{Binary Neutron Stars}
For binary neutron stars the situation is significantly different from
that for binary black holes.  Depending on the equation of state,
stellar masses and spins, the merger of binary neutron stars may be
triggered either by a plunge after the two stars reach an ISCO, or by
Roche lobe overflow (see,
e.g.~\cite{Uryu_etal_00,Taniguchi_Gourgoulhon_03}).  In either case,
the merger is expected to occur at a gravitational wave frequency of
approximately 2 kHz, outside of LIGO's most sensitive regime.
That means that the current LIGO configuration is more sensitive to
the inspiral phase, which may be well approximated by post-Newtonian 
calculations, than the merger of binary neutron stars,
which has to be modeled with numerical relativity.  Hence, 
the role of numerical relativity in observing binary
neutron stars is different from its role for binary black holes.
In the near-term, numerical simulations could be used to validate the
post-Newtonian approximation to the waveforms in the LIGO frequency
band; whether this is possible on current generation computing
facilities remains to be seen. Furthermore, numerical relativity may 
provide guidance for the design of future configurations, given the
astrophysical scenarios that seem particularly promising.  In the
long-term, the prospect of observing gravitational radiation from the merger of
binary neutron stars is very exciting because it is very rich in
physical effects that may play an important role.  Unlike
binary black holes, which are governed entirely by Einstein's field
equations, the dynamical evolution of binary neutron stars also
depends on the equation of state, magnetic fields, radiation and
neutrino transport, and possibly other effects.  Realistic, nuclear
equations of state have already been adopted in simulations of binary
neutron star mergers \cite{Shibata_Taniguchi_06}, and numerical codes
that incorporate general relativistic magnetohydrodynamics have been
developed (e.g.~\cite{Duez_etal_06,neilsenMHD}).  Detecting a binary
neutron star merger may therefore establish important observational
constraints on these aspects, in particular the equation of state.
Clearly, this is a very exciting prospect.

As discussed above, it is unlikely that the current gravitational-wave
detectors could observe the details of the merger.  Numerical
relativity may nevertheless play an important role for the purposes of
data analysis, namely by identifying features in the gravitational
wave signal that could provide particularly important information.  A
concrete example is the question whether or not the merger remnant
promptly collapses to a black hole.  Depending on the equation of
state, binary neutron stars may either collapse to a black hole on a
dynamical timescale after merger or the remnant may form a
``hypermassive'' neutron star supported against collapse by virtue of
differential rotation~\cite{Shibata_Taniguchi_06,Baumgarte_etal_00}.
For binaries with fixed masses the pre-merger gravitational-wave
signal is quite similar, but the post-merger signal differs
significantly for the two scenarios.  Distinguishing these post-merger
signals would therefore provide an important constraint on the
neutron-star equation of state.  Unfortunately, the post-merger signal
typically has a frequency close to 4 kHz \cite{Shibata_Taniguchi_06}
and may require advanced, perhaps specialized, detectors to extract
significant information (see also the discussion in
\cite{Lewis_etal_06}).  Investigations of these waves and the
information they carry could therefore be used to inform development
of advanced gravitational-wave detectors.

\subsubsection{Black hole-neutron star binaries}

For mixed black hole-neutron star binaries the situation is different
again.  The inspiral can have two very different outcomes: either the
neutron star is tidally disrupted by the black hole before it reaches
the ISCO, or it reaches the ISCO first and spirals into the black hole
more or less intact.  A very crude scaling argument suggests that the
separation of tidal disruption $s_{\rm tid}$ is approximately
\begin{equation}
\frac{s_{\rm tid}}{M_{\rm BH}} \approx 
\left( \frac{M_{\rm NS}}{M_{\rm BH}} \right)^{2/3} 
\frac{R_{\rm NS}}{M_{\rm NS}}, \label{disruption}
\end{equation}
while the ISCO is located approximately at
\begin{equation}
\frac{s_{\rm ISCO}}{M_{\rm BH}} \approx 6. \label{isco}
\end{equation}
If $s_{\rm tid} > s_{\rm ISCO}$, the neutron star is disrupted before
it reaches the ISCO, and vice-versa.  Since most neutron stars are
expected to have a mass of slightly more than a solar mass and ratio
$R_{\rm NS}/M_{\rm NS} \approx 5$, the outcome mostly depends on the
black hole mass $M_{\rm BH}$ (see Ref.~\cite{vallisneri}).

The neutron star can only be disrupted tidally outside the ISCO for
relatively small stellar-mass black holes with $M_{\rm BH} \lesssim 5
M_{\rm NS}$.  This regime is very interesting for a number of
astrophysical reasons.  Such a disruption may act as the central
engine of short gamma ray bursts, and, as for binary neutron stars, a
detection may provide useful constraints on the equation of state.  It
is also this regime that requires numerical relativity simulations for
quantitative predictions.  Extending the above crude estimates, the
orbital frequency at the onset of tidal disruption is approximately
\begin{equation} \label{omega}
\Omega_{\rm orb} \approx 
\left( \frac{M_{\rm BH}}{s_{\rm tid}^3} \right)^{1/2}
\approx \left( \frac{M_{\rm NS}}{R_{\rm NS}^3} \right)^{1/2},
\end{equation}
which is the same order as the inverse of the neutron star's
dynamical timescale.  As for the merger of binary neutron stars, the
corresponding gravitational-wave frequency is outside of LIGO's most
sensitive wave band.  The role of numerical relativity is therefore
again to explore these scenarios and identify particularly interesting
frequency regimes.

Numerical relativity simulations of mixed black hole-neutron star
binaries are not as far advanced as those of black hole binaries or
neutron star binaries.  So far, fully relativistic, self-consistent
studies exist for quasi-equilibrium models
\cite{Baumgarte_etal_04,Taniguchi_etal_06,Grandclement_06} (see also
\cite{Miller:2001zy}).  The first fully self-consistent, dynamical
simulations of the binary inspiral and the tidal disruption of the
neutron star have been announced very recently \cite{Shibata_Uryu_06}
(see also \cite{Faber_etal_06} for simulations of extreme mass-ratio
binaries within the so-called Wilson-Mathews approximation, which
assumes that the spatial metric remains conformally flat).  Other
groups have also initiated studies of mixed binaries
\cite{Bishop:2004mt,Loffler:2006wa}.  Clearly, more comprehensive
studies of tidal break-up in black hole-neutron star binaries remain
an important and urgent goal.

These calculations will help to address several very important
questions.  One such question is whether the tidal disruption of a
neutron star in a mixed binary may lead to an accretion disk that is
large enough to power a gamma ray burst.  Another question concerns
the nature of the mass transfer, which could proceed on a dynamical or
secular timescale, or could even be episodal (see, e.g., the
discussion in \cite{Faber_etal_06} and references therein).  The
nature of the tidal disruption has great impact on the gravitational
wave signal emitted in the process.  Presumably the disruption will
leave a signature in the signal at a frequency related to orbital
frequency.  As inferred from equation (\ref{omega}), this frequency
carries important information about the neutron star and its
structure.  A quantitative understanding of these issues clearly
requires detailed numerical relativity simulations.

\section{Acknowledgments}

We would like to thank
A. Buonanno, G. Cook, O. Moreschi, J. Pullin and R. Price 
for helpful discussions.  This work was
supported in part by grants from NSF: PHY-0326311
and PHY-0554793 to
Louisiana State University, PHY-0456917 to Bowdoin College,
PHY-0200852 to University of Milwaukee-Wisconsin, the CIAR, NSERC and
Alberta Ingenuity. The simulations described here were performed on
the University of British Columbia's {\bf vnp4} cluster (supported by
CFI and BCKDF), {\bf WestGrid} machines (supported by CFI, ASRI and
BCKDF), and the Dell {\bf Lonestar} cluster at the University of Texas
in Austin. L.L. thanks CIAR for support to attend the 2005 and 2006
Cosmology \& Gravity meetings where portions of this research were
started. P.R.B.  and L.L are grateful to the Alfred P Sloan Foundation and
Research Corporation for financial support.

\appendix
\section{Calculation of radiation and systematic effects} \label{appwaves}
In this section we discuss in more detail possible systematic effects
resulting in the calculation of waveforms. We concentrate in particular
in the approach based on the Newman-Penrose (spin-weighted) scalar $\Psi_4$
as is the most commonly employed, however similar issues arise in the
perturbative approach as well.

\subsection{From the analytical to the numerical arena.}

$\Psi_4$ is a particular combination of the Weyl tensor
in a suitable frame and coordinates. Its leading
order $\Psi_4^0$ in a suitable Taylor expansion off future null infinity  
(${\scri}^+$)  provides the gravitational waves. From now on, as we
concentrate on the extraction of gravitational wave we will drop
the supra-index ``${}^0$'' from all related quantities, though it must
be understood that we are referring to the leading behavior.  
A related quantity, the shear of
the outgoing null rays $\sigma$ plays a crucial role in the
calculation of radiated momentum and angular momentum.  A few key
features are trivially satisfied at ${\scri}^+$ by construction which
makes the unambiguous calculation of the radiative properties of the
spacetime possible.  Namely: the metric is exactly flat, the location
of the extraction worldtube is unique (up to time $u$ translations),
and a powerful structure (like the asymptotic transformation group,
asymptotic flatness condition and peeling behavior) allows for a clean
definition of the sought-after quantities. This ensures a unique
calculation of the radiative properties of the spacetime.

Unfortunately, most codes are unable to calculate these quantities at
future null infinity and thus $\Psi_4$ is obtained at finite
distances\footnote{Notable exceptions are provided by implementations
based on the characteristic and conformal formulations of General
Relativity (see \cite{Winicourreview} and \cite{Friedrichreview} and
references cited therein). However, these approaches have not yet been
used to tackle binary spacetimes of the types discussed here.}.  Since
the extraction worldtube does not represent a flat surface, key
ingredients are missing which can have a non-trivial impact in the
calculated quantities, even when the suitable decay of this quantity
is materialized (i.e. peeling is satisfied).  For instance, the
suitable frame allowing for the calculation of radiation at future
null infinity -- known as a Bondi frame -- is such that the angular
part of the metric is exactly that of the unit sphere (i.e. the
angular metric induced at any given time is exactly that of the unit
sphere), there is no induced `shift' in the coordinates of observers
at ${\scri}^+$ ($g_{uA}=0$) and the observers retarded-time $u$ is affinely
parametrized ($g_{ur}=0$). These conditions play a crucial role in
several aspects: (i) inertial observers maintain constant angles and
clocks that tick at a constant rate (ii) the variation of the angular
part of the metric near ${\scri}^+$ is solely due to gravitational
waves and (iii) the simple relation $\Psi_4 = {\bar \sigma}_{,uu}$
holds (with $\bar \sigma$ denoting the complex conjugate of
$\sigma$). This relation, together with the Bianchi identities at
future null infinity, is employed to replace the appearance of ${\bar
\sigma}$ by suitable integrals of $\Psi_4$.

At the numerical level, the worldtube is routinely defined by a
Cartesian timelike worldtube at $x_i^2=r^2$ intersecting hypersurfaces
at $t=\textrm{const}$. The induced metric on this worldtube generically does
not satisfy the conditions $g_{AB} = r^2 q_{AB} + O(r)$ (with $q_{AB}$
the unit sphere metric, $g_{tt}=1$ $g_{tA} = 0$. Consequently, as
discussed in detail in \cite{lehnermoreschi}, if $\Psi_4 \neq \ddot
{\bar \sigma}$ (with a ``$\; \dot{} \;$'' indicating $\partial_t$),
observer's clock rates (at different locations on the extraction
worldtube) tick at dissimilar rates and do not stay at constant
angular locations.  These issues can introduce systematic effects
which can affect the predicted outcome.  For instance,
\begin{itemize}

\item If $\Psi_4 \neq \ddot {\bar \sigma}$, commonly employed formulae
which replaces $\bar \sigma^0$ by integrals of $\Psi_4$, lack
non-trivial contributions from products of $\bar \sigma$ and (time
derivatives of) the conformal factor $F$ relating the angular metric
$g_{AB}$ to the unit sphere metric $q_{AB}$ -- see Eq.~(\ref{psi4shear}). 
Recall that since $g_{AB}$ is the metric of a
sphere, it is conformally related to that of the unit sphere by $g_{AB} =
r^2 F^2 q_{AB}$

\item If $g_{tA} \neq 0$, inertial observers suffer a rotation (from
the induced shift $\beta_A=g_{tA}$).  Therefore, predictions like the
waveforms at a particular angle will be affected as inertial
coordinates are shifting around the worldtube. This will influence the
extraction of multipole contributions, since the spin-weighted
spherical harmonics employed in such a task do not take into account
the shift in the angular coordinates (see \cite{hpgn} for a related
discussion of these issues at $\scri^+$).

\item If $g_{tt} \neq 1$ the radiation measured by different observers
at a constant $t=\textrm{const}$ slice does not correspond to the same inertial
time. Thus the extracted waveforms would have to be mapped, at each
angle to the real inertial time.

\end{itemize}

It is clear that these issues can introduce systematic effects that
can be either corrected or at least estimated within a given
simulation.  Notice that these issues can not be completely addressed
with an improved tetrad choice as any tetrad must satisfy $g_{ab} = 2
l_{(a} n_{b)} - 2 m_{(a} m_{b)}$. Thus, the induced metric in all
cases will be the same. Nevertheless a convenient choice of tetrad
aids, in particular, to alleviate issues related to the proximity of
the extraction worldtube to the source.

It is useful to first at least determine the magnitude of the 
effects the issues discussed above might have in a given simulation. Then if required, 
these effects can be  corrected by suitably modifying the
employed expressions to remove many of the ambiguities. In what follows we describe 
the first part, while a discussion of the second will be presented elsewhere\cite{lehnermoreschi}
and applied in relevant scenarios\cite{wavecorrection}.
 
First and foremost is estimating the amount that the above mentioned effects 
can have on $\Psi_4$. We will ignore here those coming due to the extraction
at finite distances as a comparison of obtained results at different radii 
can be employed to estimate this effect. Furthermore, assuming 
the extraction takes place sufficiently far away, the difference between
a derivative in the time labeling the timelike hypersurfaces and
null hypersurfaces --beyond gauge factors which we will discuss later-- 
is given by the contribution of spatial derivatives off the extraction
worldtube times $1/r$ factors due to an approximate ``potential'' from the source.
This effect is also controlled by placing the extraction sphere sufficiently far and comparing
the obtained results at different radii.
Under these assumptions, the main contribution to the error in assigning 
the radiation to the straightforwardly calculated $\Psi_4$ is given by the discrepancy
of the extraction sphere being in agreement with   that of a Bondi frame.
This is made evident by the failure of the induced metric on the sphere to be that
of the unit sphere\cite{winicour}. In turn one has $g_{AB}=r^2 S_{AB}$, which can be re-expressed
in terms of $S_{AB} = F(t,\theta,\phi)^2 q_{AB}$ with $q_{AB}$ the unit sphere metric.
The correction to  $\Psi_4$ is given by
\begin{equation}
\Psi_4^0 = \ddot{ \bar \sigma} - \bar \eth^2 \frac{\dot F}{F} + \bar \sigma \frac{ \ddot F}{F}
+ 2 \left ( \dot {\bar \sigma} \frac{\dot F}{F} - \bar \sigma \frac{(\dot F)^2}{F^2} \right ) \label{psi4shear}
\end{equation}
with $\eth$ the {\it eth} operator which is a particular combination of derivatives on the
sphere\cite{NP}.
Thus, unless $\dot F = 0$, a non-trivial correction must be considered from the fact
that the extraction sphere is not inertial. Certainly this will occur as the very
radiation one is trying to compute will be responsible for accelerating the sphere.
The key message is then to estimate the role $F$ will play. To this end two calculations
can be performed. First, a partial answer on the value of $F$ can be obtained by
simply evaluating $\tilde F =det(g_{AB})/det(r^2 q_{AB})$. If $\tilde F \neq 1$, then $F\neq 1$ and
it can therefore play a non-trivial role. However, if  $\tilde F = 1$ it is not
necessarily the case that $F=1$. A more involved, though now complete, recipe to obtain
$F$ can be easily obtained by computing the Ricci scalar associated with $g_{AB}$ and
$q_{AB}$ and the fact that the two metrics are conformally related by $F$.
This gives rise to the expression
${\cal R} = 2 \left( F^2 + \nabla^A \nabla_A \log F \right )$.
Notice that $F=1$ is a solution if $g_{AB}$ is the unit sphere metric, hence, short
of obtaining a solution for $F$, an estimate of ignoring this fact can be obtained
by ${\cal E}_F = || {\cal R} - 2 ||$. Naturally, if  ${\cal E}_F$ remains well below
the measured waveforms, the straightforward use of $\Psi_4$ would be warranted.
To simplify the numerical calculation of ${\cal R}$ one can make use of the Gauss-Codacci relations
for a 2-dimensional hypersurface $S$ in a three-dimensional manifold $\Sigma$ and employ
quantities readily available on $\Sigma$. Defining
the extrinsic curvature of $S$ by $\kappa_{ab} = h^c_a h^d_b \nabla_{c} s_d$ (with
$h_{ab} = \gamma_{ab} - s_a s_b$, the induced metric on $S$ with normal $\hat s_a=\nabla_a R$,
$s^a=\hat s^a/(\hat s_a \hat s^a)^{1/2}$ and
$\nabla_c \gamma_{ab} = 0$), a straightforward calculation indicates
\begin{equation}
{\cal R} = {}^{(3)}R - 2 \, \, {}^{(3)}R_{ab} s^a s^b - \kappa^2 + \kappa_{cd} \kappa^{cd}
\end{equation}
In addition to the conformal factor $F$ being taken properly into account, a Bondi
frame satisfies that observers measure an affine time along ${\scri}^+$ and proceed
along constant angular coordinates. These conditions will be met unless $g_{tt}=1,g_{tA}=0$.
Here again norms could be defined as ${\cal E}_{Gtt} = || (g_{tt} - 1) ||, 
{\cal E}_{GtA} = || g_{tA} ||$ so as to obtain an estimate of the effect these issues
might have in the extraction process.

\subsection{Coordinate conditions and extracted quantities---an example}

To illustrate some of the effects of coordinate conditions that are
not well adapted to the extraction mechanism we adopt a spacetime
containing linearized gravitational waves~\cite{teukolskywaves}.  For
our particular example, the spacetime is described in terms of the
following line element
\begin{eqnarray}
ds^2 &=& -dt^2 + (1+A f_{rr}) dr^2  \nonumber \\
& & + 2 B f_{r\theta} dr d\theta + 2 B f_{r\phi} \sin(\theta) dr d\phi \nonumber \\
& & + (1 + C f^1_{\theta\theta} + A f^2_{\theta\theta}) r^2 d\theta^2 \nonumber \\
& &+ 2 (A-2C)f_{\theta\phi} r^2 \sin(\theta)^2 d\theta d\phi \nonumber \\
& &+  \left (1 + C f^1_{\phi\phi}
 + A f^2_{\phi\phi} \right ) r^2 \sin(\theta)^2 d\phi^2 \label{eq:teuwaves}
\end{eqnarray}
where
\begin{eqnarray*}
f_{rr} = \sin(\theta)^2 \cos(2 \phi) \; &;& \; f_{r\theta} = \sin(2 \theta) \cos(2\phi)/2 \\
f_{r\phi} = - \sin(\theta) \sin(2\phi) \; &;& \; f_{\theta\phi} = \cos(\theta) \sin(2 \phi)\\
f^1_{\theta\theta} = -f^1_{\phi\phi} &=& (1 + \cos(\theta)^2) \cos(2\phi) \\
f^2_{\theta\theta} = -\cos(2\phi) \; ; \; f^2_{\phi\phi} &=& -\cos(\theta)^2 \cos(2\phi)
\end{eqnarray*}
and
\begin{eqnarray}
A &=& \left ( \frac{-3\sin(Y)}{r^3} + \frac{9 \cos(Y)}{r^4} + \frac{9 \sin(Y)}{r^5} \right ) \\
B &=& \left ( \frac{\cos(Y)}{r^2} + \frac{3\sin(Y)}{r^3} - \frac{6 \cos(Y) }{r^4} \right. \\
& & - \left. \frac{6 \sin(Y)}{r^5} \right ) \\
C &=& \frac{1}{4} \left (\frac{\sin(Y)}{r} - \frac{2 \cos(Y)}{r^2} - \frac{9 \sin(Y)}{r^3} \right .\\
& & + \left. \frac{21 \cos(Y) }{r^4} - \frac{21 \sin(Y)}{r^5} \right )
\end{eqnarray}
with $Y=t-r$ and for simplicity we have adopted $F=\sin(t-r)$ in
reference~\cite{teukolskywaves}.  While a detailed discussion of 
of problematic issues that can arise in the extraction proceess as well
as ways to handle them will be discussed elsewhere\cite{lehnermoreschi}
we here illustrate the effect that some of these will have in a simple scenario.

Notice that the line element \ref{eq:teuwaves} satisfies all the mentioned properties, in particular
$g_{tt}=1$, $g_{tA}=0$  and $g_{AB}=r^2 q_{AB} + O(r)$ (with $A=\theta,\phi$ and $q_{AB}=diag[1,\sin(\theta)^2]$).
For clarity, we will concentrate on two very simple cases that will violate these conditions defined by
$r\rightarrow g(t) r$ and $\phi \rightarrow \phi + \omega t$.The former introduces a time-dependent variation
in the location of the extraction radius with respect to a physically defined areal radius. The latter induces
a shift along the $\partial_{\phi}^a$ direction.
Following the commonly used approach one obtains for $\Psi_4$,
\begin{eqnarray}
\Re(\Psi_4^0) &=& \frac{  \sin(\tilde Y) (1+\cos(\theta)^2) }{4g} \cos(2 (\phi + \omega t) )  \\
\Im(\Psi_4^0) &=& \frac{- \sin(\tilde Y) \cos(\theta) }{2 g} \sin(2 (\phi + \omega t) )    
\end{eqnarray} 
with $\tilde Y = t-rg$; clearly, while a calculation of the radiated energy would be immune to the value of $\omega$
the actual waveforms would be affected. Additionally, both would be affected by the functional
dependence of $g(t)$, which even when chosen initially to be unity, it will vary as waves
propagate outwards affecting the spacetime. Clearly a generic situation would not be as simple
as the one analyzed above, though it serves to make evident how these issues can be 
obscured in a non-trivial extraction of physical quantities in an otherwise
correctly obtained numerical solution.

\section{Estimating the computational cost of binary black hole merger simulations} \label{app_cost}
Here we present an order of magnitude estimate of the computational cost
of simulating a binary black hole merger for a specified number of orbits,
and where we want to extract some feature of the solution to within a given accuracy.
More precisely, if the CPU run time and estimated net solution error of
a fiducial simulation are known, 
we give an order of magnitude estimate for the CPU run
time that a second simulation will take to obtain a new result
with the same accuracy but from an evolution that completes
a different number of orbits before merger.

Assume the numerical error $E(t,h)$ in some desired property of the 
solution (for example the phase evolution of the gravitational waveform)
has the following dependence on physical time $t$ and characteristic
discretization scale $h$:
\begin{equation}\label{ass_error}
E(t,h)\propto t^q h^m.
\end{equation}
Here $q$ is a positive constant of order unity,
and $m$ is the order of the discretization method. In general
the growth of error will be more complicated than this simple power law,
though for an order of magnitude estimate this expression is sufficient. Note
also that we have assumed the code has been ``cured'' of any exponential growth
in error.
In an adaptive code there will be several mesh spacings $h$, though the scaling
relationships derived below will still be correct if all mesh resolutions
are changed by the same factor when the resolution is changed. Equation
(\ref{ass_error}) is strictly only valid for finite-difference codes,
though in the final expression below we can take the limit as $m\rightarrow\infty$
to get estimates for spectral codes. This will give us an idea of the
scaling of present pseudo-spectral codes in the regime where the leading
source of error comes from the spatial discretization and not 
the finite-difference time stepper.

A more useful parameter describing the physical run time $t$ of the
simulation is the number of orbits $n$ completed before coalescence, and
we will assume that the inspiral regime of the merger dominates the run-time. 
We can use the leading order Post Newtonian expression for equal mass,
quasi circular inspirals to estimate $n(t)$:
\begin{equation}\label{nt}
n(t)\propto t^{5/8}
\end{equation}

The final ingredient in our scaling estimates will be the manner in which computational
run time $T$ scales with mesh spacing $h$ for any {\em optimal} grid-based solution
method of a $3+1$ dimensional system of partial differential equations
\begin{equation}\label{th}
T(h,t)\propto \frac{t}{h^4}
\end{equation}

The first question we can now answer is the following: given that simulation $A$
required $T_A$ CPU hours to complete a simulation of a binary system 
exhibiting $n_A$ orbits before coalescence, and from which we extracted
a desired quantity with error $E_A$, how long $T_B$ will it take to run second
simulation of a similar binary system, now with $n_B$ orbits 
before coalescence but the {\em same} net error $E_B=E_A$? Using 
(\ref{ass_error}-\ref{th}) we find
\begin{equation}\label{tb_ta}
T_B = T_A \left(\frac{n_B}{n_A}\right)^{(8/5 + 32q/5m)}.
\end{equation}
For example, assuming linear growth of error ($q=1$), we 
get $T_B=T_A^z$, where $z=4.8,3.2,2.7,2.4,...,1.6$ for
$m=2,4,6,8,...,\infty$ respectively. Note that the difference
going from $2^{nd}$ to $4^{th}$ order accuracy is quite
significant, as is the jump from $4^{th}$ to ``spectral'' 
convergence $m=\infty$. The same holds when estimating the
accuracy achieved by different order-of-accuracy operators 
in modeling modes in the solution \cite{ray,lehnerlieblingreula}.
This implies that while higher order methods are important, a significant 
gain is already achieved at $4^{th}$ order.

An application of the preceding expression is to the recent
survey of unequal mass inspirals presented in \cite{gonzalez_et_al}.
The $4^{th}$ order accurate code ($m=4$) discussed there is 
quite fast by today's standards, and they were
able to perform the survey utilizing a total of about
$150,000$ CPU hours (recall however that the cost of these simulations
is in practice further alleviated by exploiting the problem's symmetry). 
The majority of initial conditions ran exhibited
about $2$ orbits before merger. Suppose the survey were repeated (including
calibration runs, etc.), but now starting with initial 
conditions resulting in $4$ orbits prior to merger. Equation
(\ref{tb_ta}) suggests (assuming $q=1$) it would take
around $1.4$ million CPU hours to complete with the 
same level of overall accuracy. Early comparisons of numerical
versus PN waveforms \cite{buonanno_et_al} suggest more than
4 orbits are needed to begin to study the adequacy of various
PN approximants. Suppose $10$ were sufficient, and the survey
of \cite{gonzalez_et_al} were repeated for the purpose
of PN comparisons---(\ref{tb_ta}) then suggests $170$ million
CPU hours would be needed. This would require a $20,000$ node
cluster running continuously for 1 year. Of course, we are {\em not}
suggesting that such a survey is necessary for PN 
comparison purposes, we are merely illustrating (\ref{tb_ta}) using
actual data for the reference simulation ``$A$'' as given
in \cite{gonzalez_et_al}.

A second interesting question we can give a rough answer to is, given
simulation $A$ has accumulated an error $E_A$ for an $n_A$ orbit 
simulation, what
is the expected error for simulation $B$ that completes $n_B$ orbits,
assuming simulations $A$ and $B$ have identical resolution $h_A=h_B$?:
\begin{equation}\label{Eb_Ea}
E_B = E_A \left(\frac{n_B}{n_A}\right)^{(8q/5)}.
\end{equation}
At a first glance it might seem strange that the accumulation
of error is independent of the order of the discretization scheme.
Though observe that this does {\em not} imply that a lower order scheme is 
just as ``fast'' as a high order method even if both
methods exhibit similar growth factors $q$, for in general it will
take more resolution for a low order method to get to the
same level of error $E$ as a higher order method. From (\ref{th})
the ratio of run times $T_{h_1}/T_{h_2}$ for two different
resolutions $h_1$ and $h_2$, regardless of the order of convergence
for an optimal solution method, scales as $(h_1/h_2)^4$.

\end{document}